\documentclass[a4paper, 11pt]{article}
\usepackage[T1]{fontenc}
\usepackage{jheppub,graphicx,epsf,amssymb,amsbsy,amsfonts,amssymb,amsmath, empheq,mathrsfs,appendix,soul}
\usepackage{hyperref}

\def\be{\begin{equation}}
\def\ee{\end{equation}}
\def\bea{\begin{eqnarray}}
\def\eea{\end{eqnarray}}

\def\a{\alpha}
\def\b{\beta}
 
\def\d{\delta}
\def\e{\epsilon}
\def\f{\phi}

\def\l{\lambda}
\def\m{\mu}\def\s{\sigma}\def\l{\lambda}

\def\bg{\bar{g}}
\def\beq{\begin{eqnarray}}\def\eeq{\end{eqnarray}}
\def\ba#1\ea{\begin{align}#1\end{align}}
\def\bg#1\eg{\begin{gather}#1\end{gather}}
\def\bm#1\em{\begin{multline}#1\end{multline}}
\def\bmd#1\emd{\begin{multlined}#1\end{multlined}}

\def\a{\alpha}
\def\b{\beta}

\def\d{\delta}
\def\D{\Delta}
\def\e{\epsilon}

\def\G{\Gamma}

\def\l{\lambda}

\def\m{\mu}

\def\s{\sigma}

\def\pa{\partial}

\def\lt{\left}
\def\rt{\right}
\def\({\left(}
\def\){\right)}
\def\[{\left[}
\def\]{\right]}
\def\<{\left<}
\def\>{\right>}

\def\tn{\textnormal}

\def\a{\alpha}

\def\l{\lambda}
\def\b{\beta}

\def\d{\delta}
\def\s{\sigma} 
\def\om{\omega}

\def\m{\mu}

\def\no{\nonumber}
\def\f{\frac}

\def\D{\Delta}
\def\Om{\Omega}

\title{Celestial OPE in Self Dual Gravity}

\author{Shamik Banerjee$^{\,a,b}$, Harshal Kulkarni$^{\,c,d}$, Partha Paul$^{\,e}$}
\affiliation[a]{National Institute of Science Education and Research (NISER), Bhubaneswar 752050, Odisha, India}
\affiliation[b]{Homi Bhabha National Institute, Anushakti Nagar, Mumbai, India-400085}
\emailAdd{banerjeeshamik.phy@gmail.com}
\affiliation[c]{Department of Physical Sciences, \\ IISER Kolkata, Mohanpur, West Bengal 741246, India}
\affiliation[d]{Department of Theoretical Physics, \\ Tata Institute of Fundamental Research, Homi Bhabha Road, Mumbai 400005, India}
\emailAdd{harshalkulkarni20@gmail.com}
\affiliation[e]{Centre for High Energy Physics, Indian Institute of Science,
C.V. Raman Avenue, Bangalore 560012, India}
\emailAdd{pl.partha13@gmail.com}
\abstract{}

\abstract{In this paper we compute the celestial operator product expansion between two outgoing positive helicity gravitons in the self dual gravity. It has been shown that the self dual gravity is a $ w_{1+\infty} $-invariant theory whose scattering amplitudes are one loop exact with all positive helicity gravitons. Celestial $w_{1+\infty}$ symmetry is generated by an infinite tower of (conformally soft) gravitons which are holomorphic conserved currents. We find that at any given order only a \textit{finite} number of $w_{1+\infty}$ descendants contribute to the OPE. This is somewhat surprising because the spectrum of conformal dimensions in celestial CFT is not bounded from below. However, this is consistent with our earlier analysis based on the representation theory of $w_{1+\infty}$. The phenomenon of truncation suggests that in some (unknown) formulation the spectrum of conformal dimensions in the dual two dimensional theory can be bounded from below.}

\begin{document}
\maketitle
\flushbottom

\section{Introduction}

Celestial holography is a conjectured duality between quantum gravity in 4D asymptotically flat spacetime and a quantum field theory on the 2D celestial sphere \cite{Strominger:2017zoo,Pasterski:2016qvg, Pasterski:2017kqt}. Symmetries play an important role in this conjectured duality. The Lorentz group in 4D acts on the 2D celestial sphere as the global conformal group. So the dual theory should be a conformal field theory. Motivated by this, a new basis was introduced \cite{Pasterski:2016qvg, Pasterski:2017kqt, Banerjee:2018gce} in which the $ S $-matrix elements transform like 2D conformal correlators. Besides the two dimensional global conformal symmetry, CCFT has various infinite dimensional current algebra symmetries \cite{Sachs:1962zza, Strominger:2013jfa, He, Strominger:2014pwa, Barnich:2009se, Kapec:2016jld, Kapec:2014opa, He:2017fsb, Donnay:2020guq, Stieberger:2018onx, Banerjee:2022wht, Banerjee:2020zlg, Banerjee:2021cly, Gupta:2021cwo, Guevara:2021abz, Strominger:2021lvk, Himwich:2021dau, Ball:2021tmb, Adamo:2021lrv, Costello:2022wso, Stieberger:2022zyk, Stieberger:2023fju,Ball:2023qim,Melton:2022fsf,Banerjee:2020kaa,Costello:2022upu,Costello:2023vyy,Garner:2023izn,Donnay:2021wrk,Donnay:2022hkf}.


Operator product expansion (OPE) is a central tool to study various aspects of any CFT. In the context of celestial CFT also, OPE played an important role in identifying new symmetries \cite{Guevara:2021abz, Strominger:2021lvk, Banerjee:2020zlg}, null states \cite{Banerjee:2020zlg, Banerjee:2021cly,Banerjee:2019aoy,Banerjee:2019tam,Pano:2023slc,Banerjee:2023zip,Banerjee:2020vnt,Banerjee:2023rni,Adamo:2022wjo,Ren:2023trv,Hu:2021lrx,Ebert:2020nqf,Pasterski:2021fjn} etc. It has also found applications in the Bootstrap program \cite{Atanasov:2021cje, Ghosh:2022net}. Based on the universal singular structure of the tree-level OPE between two positive helicity gravitons, it was shown in \cite{Guevara:2021abz} that CCFT has an infinite tower of soft symmetries which close into $w_{1+\infty}$ algebra \cite{Strominger:2021lvk}. Loop corrections to the tree level celestial OPEs  have been studied in \cite{Bhardwaj:2022anh, Krishna:2023ukw}.


In a previous paper \cite{Banerjee:2023zip}, we have studied the implications of the $w_{1+\infty}$ symmetry at the level of OPEs by using representation theory. By studying the subleading terms in the OPE between two positive helicity outgoing gravitons, we have shown that there should exist an infinite number of theories which are invariant under $ w_{1+\infty} $ algebra. 


In this paper we derive the OPE in one such theory, known as the quantum self dual gravity \cite{Ooguri:1991fp, Chalmers:1996rq, Bern:1998xc, Bern:1998sv, Krasnov:2021sf} which was shown to be $ w_{1+\infty} $ invariant in \cite{Penrose:1968pr, Penrose:1976pr, Boyer:1985pn}. Here we do a collinear expansion of the known graviton scattering amplitudes in the self dual gravity theory and extract the celestial OPE from there. For simplicity, we analyse the $5$-point all plus amplitude in self dual gravity and factorize it in the collinear limit through a $4$ point amplitude. The results we obtain are consistent with what we proposed in \cite{Banerjee:2023zip} based on the representtion theory of $w_{1+\infty}$. The rest of the paper is organised as follows. 

In section \ref{nac} we introduce notations and conventions used in this paper. Section \ref{reviewalg} briefly describes the $ w_{1+\infty} $ algebra and how the whole tower of the $ w_{1+\infty} $-currents can be generated using the two $ sl_2(R) $ sub-algebras. In section \ref{sasd} we briefly discuss about the scattering amplitudes in quantum self dual gravity. Section \ref{OPEA} discusses how to extract the OPE between two positive helicity outgoing gravitons from the 5-point one-loop self dual amplitude. We start by simplifying the 5-point amplitude in the momentum space and then Mellin transform it to get the celestial amplitudes. We then discuss how to factorize each term order by order in the OPE limit of the celestial amplitudes. The null states of the self dual gravity appearing at various orders of the OPE and the invariance of the OPE under $ w_{1+\infty} $ algebra are discussed in Appendices \ref{nssd} and \ref{invsd}. 

For the sake of completeness of the paper, we give a brief review of the celestial amplitude in Appendix \ref{review}. In Appendix \ref{dp}, we discuss the parametrization of the $ 4 $ and $ 5 $-point delta functions which are useful in our context of the OPE expansion. Appendices \ref{4pt_sd_mom_simp} and \ref{5pt_sd_mom_simp} discuss how to simplify the $ 4 $ and $ 5 $-point amplitudes in momentum space using momentum conserving delta functions and various identities of the spinor-helicity brackets. These simplifications are done keeping in mind the fact that we want to factorize the $ 5 $-point amplitude in terms of the $ 4 $-point amplitude in the OPE expansion. Appendix \ref{mellin_5_pt} deals with the Mellin transformation of the $ 5 $-point amplitude. In Appendix \ref{wap} we discuss the conditions on the graviton primary operators under the $ w_{1+\infty} $ algebra which follow from the universal structure of the OPE. In Appendix \ref{mnt}, we list the transformation properties of all the MHV null states under the action of $ {sl_2(R)}_V $ and $ \overline{sl_2(R)} $ algebras which are required to show the $ w_{1+\infty} $-invariance of the self dual OPE. Appendix \ref{revgs} briefly reviews the construction of a general $ w_{1+\infty} $-algebra invariant OPE and how one can obtain an infinite family of $ w_{1+\infty} $-algebra invariant theories.
  
\section{Notations and Conventions}
\label{nac}

In this paper, we will work in the $ (2,2) $ signature space-time, which is also known as Klein space. The null momentum $ p^\mu $ of a massless particle, satisfying the onshell condition $  p^2 = 0 $, is parametrized as,
\be \label{mom}
\begin{split}
p^\mu &= \e \, q^\mu \\
q^\mu &= \om \{1+z \bar z , z+\bar z, z-\bar z, 1-z \bar z\} 
\end{split}
\ee
where $ \e=\pm 1 $ for outgoing and incoming particles respectively, $ (z, \bar z) $ are two independent real variables and $ \om $ is any positive number interpreted as the energy of the particle. In Klein space the null infinity takes the form of a Lorentzian torus (known as the celestial torus) times a null line. The Lorentz group in $ (2,2) $ signature is given by $ SO(2,2) \simeq \f{SL(2,\mathbb{R})_L \times SL(2,\mathbb{R})_R}{\mathbb{Z}_2} $ and acts as the group of conformal transformations on the celestial torus:
\be  \label{lt}
\begin{gathered}
SL(2,\mathbb{R})_L: \qquad z \to \f{a z + b}{cz+d}, \ \bar z \to \bar z, \ ad-bc=1, \\
SL(2,\mathbb{R})_R: \qquad \bar z \to \f{\bar a \bar z + \bar b}{\bar c \bar z + \bar d}, \ z \to z, \ \bar a \bar d - \bar b \bar c = 1.
\end{gathered}
\ee
In our conventions the spinor-helicity variables are given by,
\be
\<ij\> = 2\e_i\e_j \sqrt{\om_i\om_j} z_{ij}, \ [ij] = 2\sqrt{\om_i\om_j} \bar z_{ij} 
\ee
where $ z_{ij} = z_{i} - z_{j} $ and we also have $ 2 p_i\cdot p_j = - \<ij\>  [ij]$.

\section{Review of $ w_{1+ \infty} $ Algebra}
\label{reviewalg}

We start by reviewing the $ w_{1+ \infty} $ algebra which follows from the universal singular terms in the OPE between two positive helicity outgoing gravitons. Let $ G^{+}_{\D}(z,\bar z) $ denote the positive helicity graviton conformal primary operator of dimension $ \D $ at the point $ (z,\bar z) $ on the celestial torus. The universal singular terms in the OPE are given by
\be\label{uni}
\begin{gathered}
G^{+}_{\D_1}(z_1,\bar z_1) G^+_{\D_2}(z_2,\bar z_2) = -\frac{\bar z_{12}}{z_{12}} \sum_{n=0}^{\infty} B\(\D_1 -1 +n, \D_2 -1\) \frac{{\bar z_{12}}^n}{n!} \bar\partial^n G^+_{\D_1+\D_2}(z_2,\bar z_2) 
\end{gathered}
\ee
  
Let us define an infinite family of positive helicity conformally soft \cite{Donnay:2018neh, Pate:2019mfs, Fan:2019emx, Nandan:2019jas, Adamo:2019ipt, Puhm:2019zbl, Guevara:2019ypd} gravitons \cite{Guevara:2021abz} as, 
\be
H^k(z,\bar z) = \lim_{\D \rightarrow k} (\D-k) G^+_{\D}(z,\bar z), \ k =1,0,-1,-2,...
\ee
with weights $\(\frac{k+2}{2}, \frac{k-2}{2}\)$. It follows from the OPE \eqref{uni} that we can introduce the following truncated mode expansion
\be\label{mode1}
H^k(z,\bar z) = \sum_{m = \frac{k-2}{2}}^{\frac{2-k}{2}} \frac{H^k_m(z)}{\bar z^{m + \frac{k-2}{2}}}
\ee
and the modes $H^k_m(z)$ are the conserved holomorphic currents. The currents $H^k_m(z)$ can be further mode expanded in the $ z $-variable to get,
\be\label{mode2}
H^k_m(z) = \sum_{\alpha \in \mathbb{Z} - \frac{k+2}{2}} \frac{H^k_{\alpha,m}}{z^{\alpha+ \frac{k+2}{2}}}
\ee
and one can show \cite{Guevara:2021abz} that the modes $H^k_{\alpha,m}$ satisfy the algebra\footnote{Here we are assuming that $\kappa=\sqrt{32\pi G_N}=2$.}
\be\label{hsa}
\begin{gathered}
\[ H^k_{\alpha,m}, H^l_{\beta,n}\] \\ = - \[ n(2-k) - m(2-l)\] \frac{\( \frac{2-k}{2} -m + \frac{2-l}{2} - n -1\)!}{\( \frac{2-k}{2} -m \)! \( \frac{2-l}{2} - n\)!}\frac{\( \frac{2-k}{2} +m + \frac{2-l}{2} +n -1\)!}{\( \frac{2-k}{2} +m \)! \( \frac{2-l}{2} +n\)!} H^{k+l}_{\alpha+\beta, m+n}
\end{gathered}
\ee
This is called the Holographic Symmetry Algebra (HSA). Now if we make the following redefinition (or discrete light transformation)\cite{Strominger:2021lvk}
\be
w^p_{\alpha, m} = \frac{1}{2}\(p-m-1\)! \(p+m-1\)! H^{-2p+4}_{\alpha,m}
\ee
then \eqref{hsa} turns into the $w_{1+\infty}$ algebra\footnote{This is the wedge subalgebra of $w_{1+\infty}$.}
\be\label{w}
\[w^p_{\alpha,m}, w^q_{\beta,n}\] = \[ m(q-1) - n(p-1)\] w^{p+q-2}_{\alpha+\beta,m+n}
\ee
For our purpose it is more convenient to work with the HSA \eqref{hsa} rather than the $w_{1+\infty}$ algebra. However, we continue to refer to the HSA as the $w$ algebra. 

\begin{figure}[h!]
\centering
\includegraphics[height=8.5cm, width=8.5cm]{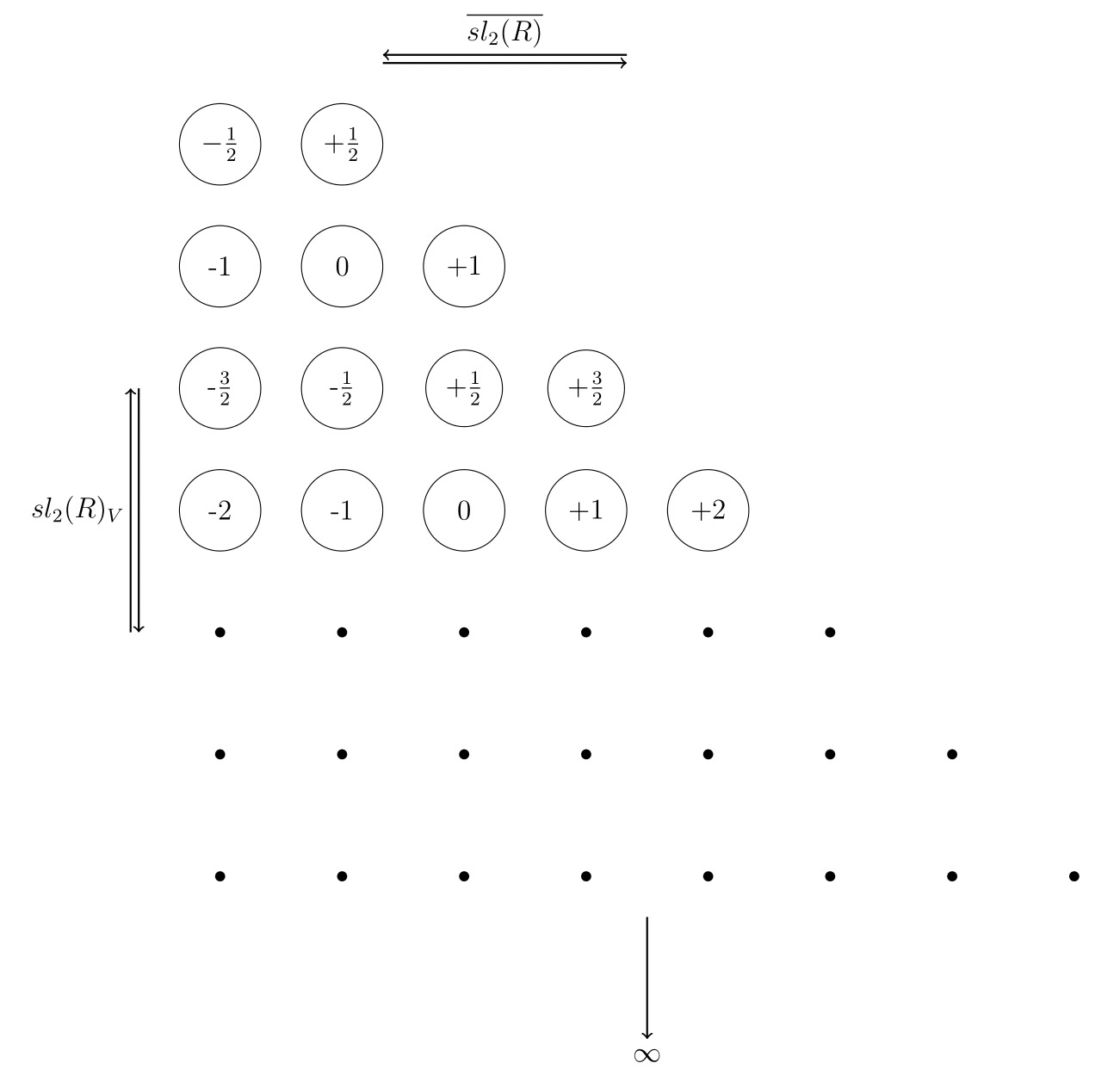}
\caption{The figure shows the soft currents. The rows and the columns are indexed by the $\overline{sl_2(R)}$ weights and the dimension $(\D=k=1,0,-1,-2,...)$ of the conformally soft graviton $H^k(z,\bar z)$, which generates the currents sitting in a row, respectively. $\overline{sl_2(R)}$ acts horizontally along a row and $sl_2(R)_V$ acts vertically along a column. In this way they generate the whole symmetry algebra starting from the current $H^1_{\frac{1}{2}}(z)$ on the top left corner.}
\label{fig}
\end{figure}

Now, in \cite{Banerjee:2023zip}, it was shown that the whole tower of the $ w $-currents can be generated using the two $ sl_2(R) $ sub-algebras. One of them is $sl_2(R)_V$\footnote{Here V stands for vertical. Please see Fig.\ref{fig} for an explanation. } generated by the operators $\{H^1_{-\f{1}{2},-\f{1}{2}}, \ H^0_{0,0}, \ H^{-1}_{\f{1}{2},\f{1}{2}}\}$, 
\be\label{vsl}
\begin{gathered}
\[ H^0_{0,0}, H^1_{-\f{1}{2},-\f{1}{2}}\] = H^1_{-\f{1}{2},-\f{1}{2}} \\
\[ H^0_{0,0}, H^{-1}_{\f{1}{2},\f{1}{2}}\] = - H^{-1}_{\f{1}{2},\f{1}{2}} \\
\[H^1_{-\f{1}{2},-\f{1}{2}}, H^{-1}_{\f{1}{2},\f{1}{2}}\] = - H^0_{0,0}
\end{gathered}
\ee
The other $sl_2(R)$ sub-algebra is generated by the global (Lorentz) conformal transformations $\{ H^0_{0,1},H^0_{0,0},H^0_{0,-1}\}$. We call this $\overline{sl_2(R)}$ because this acts only on the $\bar z$ coordinate. Now the $w$ symmetry is generated by the infinite number of soft currents $\{H^k_{p}(z)\}$ where $k=1,0,-1,-2,...$ is the dimension $(\D)$ of the soft operator and $\frac{k-2}{2}\le p\le - \frac{k-2}{2}$. For a fixed $k$, the soft currents $\{H^k_{-\frac{k-2}{2}}(z),...,H^k_{\frac{k-2}{2}}(z)\}$ transform in a spin $\frac{2-k}{2}$ representation of the $\overline{sl_2(R)}$. 

Now let us consider the currents $\{H^1_{\frac{1}{2}}, H^0_{1},...,H^k_{\frac{2-k}{2}},...\}$ with the \textit{lowest} $\overline{sl_2(R)}$ weights. These currents transform in an irreducible highest weight representation of the $sl_2(R)_V$. This can be seen from the following commutation relations following from \eqref{hsa},
\be
\begin{gathered}
\[ H^{-1}_{\frac{1}{2},\frac{1}{2}}, H^k_{\alpha,\frac{2-k}{2}}\] = - \frac{1}{2}(k-2)(k-3) H^{k-1}_{\alpha+\frac{1}{2},\frac{2-(k-1)}{2}}\\
\[ H^{0}_{0,0}, H^k_{\alpha,\frac{2-k}{2}}\] = (k-2) H^{k}_{\alpha,\frac{2-k}{2}}\\
\[ H^{1}_{-\frac{1}{2},-\frac{1}{2}}, H^k_{\alpha,\frac{2-k}{2}}\] = - H^{k+1}_{\alpha-\frac{1}{2},\frac{2-(k+1)}{2}}
\end{gathered}
\ee
Therefore, starting from the current $H^1_{\frac{1}{2}}(z)$ we can generate \textit{any} other $w$ current by the combined action of the $\overline{sl_2(R)}$ and $sl_2(R)_V$ (Fig.\ref{fig}). 

\section{Scattering Amplitudes in Quantum Self Dual Gravity}
\label{sasd}

In this section, following \cite{Bern:1998xc} we briefly review the all-plus helicity scattering amplitudes in quantum self dual gravity. In $ (2,2) $ signature, self duality translates into the following condition on the Riemann tensor,
\be 
R_{\mu\nu\rho\s} = \f{1}{2}{\varepsilon_{\m\nu}}^{\a\b}R_{\a\b\rho\s}
\label{selfd_cond}
\ee
where $ \varepsilon^{\m\nu\a\b} $ is the completely antisymmetric tensor with $ \varepsilon^{0123} = +1 $. In order to maintain the reality condition on the fields, the self dual gravity is described in either $ (2,2) $ or $ (0,4) $ signature. In Lorentzian $ (1,3) $ signature, the condition \eqref{selfd_cond} acquires an extra factor of $ i $ and contain no real solutions.  At the classical level, the linearized self dual solutions consist of positive helicity plane waves. 

In this paper, we are interested in the collinear behaviour of gravitons in the self dual gravity. At the tree level, we have only one non-trivial amplitude: the three point $\overline{MHV}$ amplitude where, only one external graviton has negative helicity and other two have positive helicity. The appearance of the negative helicity graviton in the three point $\overline{MHV}$ amplitude can be explained from the fact that the action contains a Lagrange multiplier which is physically interpreted as the negative helicity graviton.

At the one loop level, the only non-zero amplitudes are the ones with all plus helicity gravitons with the minimum number of gravitons being four. These amplitudes are both UV and IR finite. The only divergences of these amplitudes are collinear and soft divergences. Our interest in this theory stems from the fact that this is a non-trivial quantum theory which is known to be $w$ invariant. 

The one loop all-plus $ n $-graviton stripped amplitude in self dual gravity is given by \cite{Bern:1998xc},
\bea
\no A_n(1^+,2^+,
...,n^+)=-\f{i}{(4\pi)^2 960} \(-\f{\kappa}{2}\)^n \sum_{\substack{1\leq a <b \leq n \\ M,N}} h(a,M,b) h(b,N,a) tr^3[aMbN]\\
\label{npt_sdg}
\eea
where $a$ and $b$ are the external legs and $M$ and $N$ are two sets such that $M \bigcup N = {1,...a-1,a+1,...b-1,b+1,...n}$ and $M \bigcap N = \phi$. The sum is over all possible $(a,b)$ and $(M,N)$, where $(M,N)$ and $(N,M)$ are not distinguished. The trace is defined as,
\be 
tr[aMbN] = \< a|K_M|b\]\<b|K_N|a\] + \[ a|K_M|b\>\[b|K_N|a\>
\ee
where $ K_M=\sum_{i\in M} k_i  $. The ``half-soft'' function $h$ is given by,
\be
\begin{gathered}
h(a,\{1,2,\cdots,n\},b) = \f{[12]}{\<12\>}\f{\<a|K_{1,2}|3\] \<a|K_{1,3}|4\] \cdots \<a|K_{1,n-1}|n\] }{\<23\>\<34\>\cdots \<n-1, n\> \<a1\> \<a2\> \cdots \<an\> \<1b\>\<nb\>} + \mathcal{P}(2,3,\cdots, n)
\end{gathered} 
\ee
where $ K_{1,m}=\sum_{i=1}^m k_i $ and $ \mathcal{P}(2,3,\cdots, n) $ represents all permutations keeping the first leg fixed. Throughout this paper we will set $ \kappa=2 $.

\section{Graviton-Graviton OPE from Self Dual Amplitudes}
\label{OPEA}

In this section, we take the $4$ and $5$ point all plus amplitudes and express them in the conformal primary basis by (modified) Mellin transformation. Then we take the (collinear) OPE limit $ (z_{45}\to 0, \bar z_{45} \to 0) $ in the 5-point amplitude with the aim of factorizing it into some differential operators acting on the 4-point amplitude at every order in the $(z_{45},\bar z_{45})$ expansion. Let us now closely look at the 4-point amplitude, first in momentum space and then in Mellin space.\footnote{For the sake of convenience of the reader we have moved some of the intermediate steps in the calculations to an Appendix. We have refereed to the Appendix in the main text whenever necessary.}

\subsection{4-Point Momentum Space Amplitude}
From \eqref{npt_sdg}, the 4-point amplitude is given by,
\bea
 A_4(1^+,2^+,3^+,4^+) &=& -\f{i}{(4\pi)^2 960}\ B_4
\label{4pt_sdg}
\eea
where
\be\label{frpt}
B_4 = \sum_{\substack{1\leq a <b \leq 4 \\ M,N}} h(a,M,b) h(b,N,a) tr^3[aMbN]
\ee
Using the explicit expressions for the trace and the ``half-soft'' functions, $ B_4 $ can be easily evaluated and then simplified to get (see Appendix \ref{4pt_sd_mom_simp} for details),
\bea
B_4  = - 2^4  \left[ \f{\lt< 13\rt> \lt< 23 \rt> ([13][23])^3}{\left< 1 5\right>^2 \left< 25\right>^2} + \(2 \leftrightarrow 3\) + \(1 \leftrightarrow 3\) \]
\eea
where we have relabelled 4 as 5. In terms of $ (\om, z, \bar z) $ variables, the above equation becomes
\bea
\no B_4 = -2^8 \left[ \e_1\e_2\f{\om_1 \om_2 \om_3^4}{\om_5^2}\f{z_{13} z_{23} (\bar z_{13} \bar z_{23})^3}{z_{1 5}^2 z_{25}^2} + \(2 \leftrightarrow 3\) + \(1 \leftrightarrow 3\)\right]
\eea
This is the form of the 4-point momentum space amplitude that we use in evaluating the Mellin transform and other manipulations.

\subsection{4-Point Mellin Amplitude}
The modified Mellin transform of the $ n $-point amplitude is given by,
\be\label{mellinmodm}
\mathcal M_n\big(\{u_i,z_i, \bar z_i, h_i, \bar h_i\}\big) = \prod_{i=1}^{n} \int_{0}^{\infty} d\omega_i \ \omega_i^{\D_i -1} e^{-i\sum_{i=1}^n \epsilon_i \omega_i u_i} A_n\big(\{\omega_i,z_i,\bar z_i, \sigma_i\}\big)
\ee 
where $u$ can be thought of as a time coordinate and $\epsilon_i = \pm 1$ for an outgoing (incoming) particle. Note that $  A_n\big(\{\omega_i,z_i,\bar z_i, \sigma_i\}\big) $ in \eqref{mellinmodm} is the full momentum space amplitude including the momentum conserving delta function.  Using \eqref{mellinmodm} we now Mellin transform the $ 4 $-point momentum space amplitude \eqref{4pt_sdg}. Using the parametrization of 4-point delta function given by \eqref{4pt_delta}, we get the full 4-point Mellin amplitude as
\be 
\begin{gathered}
 \mathcal{M}_4(1^+_{\D_1},2^+_{\D_2},3^+_{\D_3},5^+_{\D_5}) = \f{i}{(4\pi)^2 960}2^6 \f{\G(\D')}{(i\mathcal{D})^{\D'}}\d(x-\bar x) \prod_{k=1}^3 \(\e_k\s_{k,1}\)^{\D_k-1} \\
\times \[ \mathcal{N}_4 + \mathcal{N}_4\(1 \leftrightarrow 3\) + \mathcal{N}_4\(2 \leftrightarrow 3\) \] 
\end{gathered}
\label{4pt_mellin}
\ee

where $ \D' = \D_1+\D_2+\D_3+\D_5 $ and  
\be \label{defN_4}
\begin{split}
\mathcal{N}_4 &= \s_{1,1}\, \s_{2,1}\, \s_{3,1}^{4}\f{z_{13} z_{23} (\bar z_{13} \bar z_{23})^3}{z_{1 5}^2 z_{25}^2} \\
\mathcal{D} &= \sum_{k=1}^3 \s_{k,1}u_{k5}
\end{split}
\ee
$\mathcal{N}_4 (1 \leftrightarrow 3)$ and $\mathcal{N}_4 (2 \leftrightarrow 3)$ corresponds to $\mathcal{N}_4$ with the points (1,3) and (2,3) interchanged, respectively. The expressions for $ \s_{i,j} $ are given in Appendix \ref{dp}. Note that when we interchange the points $ (1,2,3) $ in $\mathcal{N}_4 $, only the first subscript in $ \s_{i,j} $ changes, second one remains unchanged.

\subsection{5-point Amplitude in Self Dual Gravity}
\label{5pt_sd_mom}

The 5-point one loop all plus helicity stripped amplitude (without the momentum conservation delta function) is given by
\bea
 A_5(1^+,2^+,3^+,4^+,5^+)=\f{i}{(4\pi)^2 960} B_5
\label{5pt_sdg_b5}
\eea
where 
\bea
B_5 &=&  \sum_{\substack{1\leq a <b \leq 5 \\ M,N}} h(a,M,b) h(b,N,a) tr^3[aMbN]
\label{5pt_sdg_ly_b5}
\eea
The above expression consists of 30 distinct terms in total. The expression of $B_5$ has been explicitly computed and simplified in the Appendix \ref{5pt_sd_mom_simp}. Its simplified form gives:
\bea
\no B_5 = -8 \left[ \f{[25]\left< 13 \right>\left< 34 \right> ([13][34])^3}{\lt<25\rt>\left< 12 \right> \left< 15 \right> \left< 24 \right>\left< 54 \right>} + \f{[24]\left< 13 \right>\left< 35 \right>([13] [35])^3}{\lt<24\rt>\left< 12 \right> \left< 14 \right> \left< 25 \right>\left< 45 \right>} + \f{[15]\left< 23 \right>\left< 34 \right>([23] [34])^3}{\lt<15\rt>\left< 21 \right> \left< 25 \right> \left< 14 \right>\left< 54 \right>} \right. \\
\no \left. \f{[14]\left< 23 \right>\left< 35 \right>([23] [35])^3}{\lt<14\rt>\left< 21 \right> \left< 24 \right> \left< 15 \right>\left< 45 \right>} + \f{[45]\left< 13 \right>\left< 23 \right>([13] [23])^3}{\lt<45\rt>\left< 14 \right> \left< 15 \right> \left< 42 \right>\left< 52 \right>} + \f{[34]\left< 25 \right>\left< 15 \right>([15] [25])^3}{\lt<34\rt>\left< 13 \right> \left< 14 \right> \left< 32 \right>\left< 42 \right>} \right. \\
\no \left. \f{[35]\left< 14 \right>\left< 24 \right>([14] [24])^3}{\lt<35\rt>\left< 13 \right> \left< 15 \right> \left< 32 \right>\left< 52 \right>} + \f{[12]\left< 34 \right>\left< 35 \right>([34] [35])^3}{\lt<12\rt>\left< 41 \right> \left< 42 \right> \left< 15 \right>\left< 25 \right>} + \f{[12]\left< 35 \right>\left< 45 \right>([35] [45])^3}{\lt<12\rt>\left< 31 \right> \left< 32 \right> \left< 14 \right>\left< 24 \right>} \right. \\
\no \left. \f{[12]\left< 34 \right>\left< 45 \right>([34] [45])^3}{\lt<12\rt>\left< 31 \right> \left< 32 \right> \left< 15 \right>\left< 25 \right>} \right] + \( 1 \leftrightarrow 3 \) + \( 2 \leftrightarrow 3 \) \\
\label{5pa}
\eea
To avoid complication, we will not write down the Mellin transformation for the full 5-point amplitude. Rather, we will first expand the 5-point amplitude around $ z_{45} = 0, \ \bar{z}_{45}=0$ in momentum space and then Mellin transform the individual terms in that expansion. 

\subsection{Expansion of the 5-Point Amplitude around $ z_{45} = \bar{z}_{45} = 0 $ in Momentum Space}
\label{ope_decomp_5pt}
 By parameterizing \eqref{5pa} in terms of $ \{\om, z, \bar z \} $ one may think that there are holomorphic singularities in the limit $ z_4 \to z_5 $ which goes like $ \f{1}{z_{45}} $. But this is not true. Clubbing together all the twelve singular-looking terms, and rewriting them gives contributions only at leading $\mathcal{O}(\frac{\bar{z}_{45}}{z_{45}})$ and higher orders (see Appendix \ref{5pt_sd_mom_simp} for details). By parameterizing $ \om_4 = t\, \om_P, \ \om_5 = (1-t)\om_P  $ we arrange all the terms in \eqref{5pa} in the following way:
\be
B_5 = -2^7 \f{\om_P}{t(1-t)} \(\f{\bar z_{45}}{z_{45}} T_L + T_{\mathcal{O}(1)} + \bar z_{45} T_{\bar z}\) + \textnormal{Higher Order Terms}
\label{b5_expanded}
\ee
where
\be
\begin{gathered}
T_L = \[\e_1 \f{z_{12}z_{25}\bar z_{12}^3 \bar z_{25}^3}{z_{13}^2 z_{35}^2}\f{\om_1 \om_2^4 }{\om_3^2} + \e_2 \f{z_{12}z_{15}\bar z_{12}^3 \bar z_{15}^3}{z_{23}^2 z_{35}^2}\f{\om_1^4\om_2 }{\om_3^2} + \e_3 \f{z_{13}z_{15}\bar z_{13}^3 \bar z_{15}^3}{z_{23}^2 z_{25}^2}\f{\om_1^4\om_3 }{\om_2^2}\]\\
+ \[\e_1 \e_2 \f{z_{13}z_{23}\bar z_{13}^3 \bar z_{23}^3}{z_{15}^2 z_{25}^2} \frac{\om_1 \om_2 \om_3^4}{\om_P^3} + \(1 \leftrightarrow 3\) + \( 2\leftrightarrow 3 \) \]
\end{gathered} 
\ee
The expressions for $ T_{\mathcal{O}(1)}, \ T_{\bar z} $ and the detailed calculation about how we arrived at these expressions are given in the Appendix \ref{5pt_sd_mom_simp}. The point we want to emphasize here is that, \eqref{b5_expanded} is the expansion of the 5-point amplitude around $ z_{45} = \bar{z}_{45}=0 $  in the momentum space. One should not confuse it with the OPE expansion on the celestial torus, which will be done in the following subsections. The terms $ T_L, T_{\mathcal{O}(1)}, \ T_{\bar z} $ contain energy factors $ \{\om_1,\om_2,\om_3\} $ which will contribute to the OPE expansion after Mellin transformation. On top of that we have 5-point momentum conserving delta functions as well as other factors in the Mellin integral, all of which will contribute in the OPE limit of the 5-point Mellin amplitude. \eqref{b5_expanded} is just a neat way of organizing the 5-point momentum space amplitude, which allows us to easily extract the OPE from the 5-point celestial amplitude.

\subsection{Mellin Transformation of the 5-Point Amplitude and Extracting the Graviton-Graviton OPE}
\label{opext}
Let us start with the modified Mellin transformation of $B_5$ given by:
\bea
\widetilde{B}_5 = \int_0^\infty \prod_{i=1}^5 d\om_i \, \om_i^{\D_i-1} e^{-i\sum_{i=1}^5 \e_i \om_i u_i} B_5 \d^{(4)}\(\sum_{i=1}^5 \e_i \om_i q_i \)\label{mellin_trans_b5}
\eea
In the above equation for $ B_5 $, we use the expansion \eqref{b5_expanded}. Then using the 5-point delta function parametrization given in the Appendix \ref{fpd}, we can extract each term in the OPE factorization in the Mellin space. We now discuss the terms order by order in the OPE expansion in Mellin space.

\subsubsection{Leading Order}
For convenience let's take $ \e_4=\e_5=+1 $. Then the leading order term in \eqref{mellin_trans_b5} is given by
\begin{align}
\no \tilde{B}_5|_{\mathcal{O}(\frac{\bar{z}_{45}}{z_{45}})} 
&= -2^6 \f{\bar z_{45}}{z_{45}}B(\D_4-1,\D_5-1)\f{\G(\D)}{\(i\mathcal{D}\)^{\D}}\d(x-\bar x) \\ &\times \prod_{k=1}^3 (\e_k\s_{k,1})^{\D_k-1} \[ \mathcal{N}_4 + \mathcal{N}_4\(2\leftrightarrow 3\) + \mathcal{N}_4\(1\leftrightarrow 3\) \] \label{lead_B5}
\end{align}
where $ \D=\sum_{i=1}^5 \D_i $. This has been derived in detail in Appendix \ref{mellin_5_pt}. Finally, taking care of the pre-factors, we can write down the Mellin transformation of the complete 5-point amplitude $A_5$ \eqref{5pt_sdg_b5} at $\mathcal{O}(\frac{\bar{z}_{45}}{z_{45}})$:
\be 
\begin{gathered}
\mathcal{M}_5|_{\mathcal{O}(\frac{\bar{z}_{45}}{z_{45}})} = -\f{i}{(4\pi)^2 960} 2^6 B(\D_4-1,\D_5-1)\f{\G(\D)}{\(i\mathcal{D}\)^{\D}}\d(x-\bar x) \f{\bar z_{45}}{z_{45}} \prod_{k=1}^3 (\e_k\s_{k,1})^{\D_k-1} \\
\times \[ \mathcal{N}_4 + \mathcal{N}_4\(2\leftrightarrow 3\) + \mathcal{N}_4\(1\leftrightarrow 3\) \]
\end{gathered}
\label{lead_ord_mellin_amp}
\ee
This gives us the 5-point Mellin amplitude at leading order. In terms of the 4-point Mellin amplitude $\mathcal{M}_4(1^+_{\D_1}, 2^+_{\D_2},3^+_{\D_3},5^+_{\D_4+\D_5}) $ given by \eqref{4pt_mellin}, we can write \eqref{lead_ord_mellin_amp} as follows: 
\be 
\begin{gathered}
\mathcal{M}_5(1^+_{\D_1}, 2^+_{\D_2},3^+_{\D_3},4^+_{\D_4},5^+_{\D_5}) = - \frac{\bar{z}_{45}}{z_{45}} B(\Delta_4 - 1, \Delta_5 - 1) \mathcal{M}_4(1^+_{\D_1}, 2^+_{\D_2},3^+_{\D_3},5^+_{\D_4+\D_5}) + \cdots 
\end{gathered}
\ee
Thus, at the level of OPE we have
\be
\boxed{
G^{+}_{\D_4}(z_4,\bar z_4) G^{+}_{\D_5}(z_5,\bar z_5) =  - \frac{\bar{z}_{45}}{z_{45}} B(\Delta_4 - 1, \Delta_5 - 1) G^{+}_{\D_4+\D_5}(z_5,\bar z_5)  + \cdots
}
\ee

\subsubsection{$  \mathcal{O}(1)$ Terms} 
\label{sec:O1}
Now we turn our attention to the $\mathcal{O}(1)$ terms in the five point Mellin amplitude. The complete expression for the 5-point Mellin amplitude at $\mathcal{O}(1)$ is given by \eqref{ord_1_app_2},
\be\label{order_1_terms}
\begin{gathered}
 \mathcal{M}_5\big|_{\mathcal{O}(1)} = - \f{i}{(4\pi)^2 960}2^5 \f{ \G(\D)}{\left( i \mathcal{D} \right)^\D} \prod_{i=1}^3 (\e_i \s_{i,1})^{\D_i-1} \sum_{k=0}^4 B(\D_4+k-1,\D_5-1) \mathcal{F}^{(1)}_{k}(\{\e_i, z_i,\bar z_i\}) \d\( x - \bar x  \)
\end{gathered} 
\ee
where $ \mathcal{F}^{(1)}_{k}(\{\e_i, z_i,\bar z_i\}) $ are some functions of its arguments, but their explicit expressions are not important for OPE factorizations. Now we take the leading conformal soft limit $ {\D_4 \to 1} $ in the above equation to get:
\be 
\begin{gathered}
\lim_{\D_4 \to 1} (\D_4-1) {\mathcal{M}}_5\big|_{\mathcal{O}(1)} = - \f{i}{(4\pi)^2 960} 2^5 \f{\G\(\sum_{i=1,i\neq 4}^5\D_i+1\)}{\(i\mathcal{D}\)^{\sum_{i=1,i\neq 4}^5\D_i+1}}\prod_{i=1}^3 (\e_i\s_{i,1})^{i\l_i} \mathcal{F}^{(1)}_{0}(\{\e_i, z_i,\bar z_i\}) \d(x-\bar x)
\end{gathered}
\label{eq:317}
\ee
Only the $k = 0$ term in the sum in \eqref{order_1_terms} survives because, in the $\Delta_4 \to 1$ limit, $B(\D_4+k-1,\D_5-1)$ is non-singular for all $k > 0$. \\
\\
On the other hand, from the leading soft graviton theorem we know that
\be 
\begin{gathered}
\lim_{\D_4 \to 1} (\D_4-1) {\mathcal{M}}_5\big|_{\mathcal{O}(1)} = \mathcal{H}^1_{-\f{3}{2},\f{1}{2}}(5) \mathcal{M}_4(1^+_{\D_1},2^+_{\D_2},3^+_{\D_3},5^+_{\D_5})
\end{gathered}
\label{eq:318}
\ee
To make things transparent, we have used $ \mathcal{H} $-notations when the soft modes are acting on the Mellin amplitudes as differential operators and the number $ 5 $ in the argument of $ \mathcal{H} $ denotes that it's a descendant of the $ 5 $-th conformal graviton primary. The consistency of the two equations \eqref{eq:317} and \eqref{eq:318} implies that:
\be 
\begin{gathered}
- \f{i}{(4\pi)^2 960} 2^5 \f{\G\(\sum_{k=1,k\neq 4}^5\D_k+1\)}{\(i\mathcal{D}\)^{\sum_{k=1,k\neq 4}^5\D_k+1}}\prod_{k=1}^3 (\e_k\s_{k,1})^{i\l_k} \mathcal{F}^{(1)}_{0}(\{\e_i, z_i,\bar z_i\}) \d(x-\bar x) \\
=  \mathcal{H}^1_{-\f{3}{2},\f{1}{2}}(5) \mathcal{M}_4(1^+_{\D_1},2^+_{\D_2},3^+_{\D_3},5^+_{\D_5})
\end{gathered}\label{eq:319}
\ee
Now, we can replace $ \D_5 $ by $ \D_4+\D_5-1 $ in \eqref{eq:319} and then use it in \eqref{order_1_terms} to get: 
\be 
\begin{gathered}
\mathcal{M}_5\big|_{\mathcal{O}(1)} = B(\D_4-1,\D_5-1) \mathcal{H}^1_{-\f{3}{2},\f{1}{2}}(5) \mathcal{M}_4(1^+_{\D_1},2^+_{\D_2},3^+_{\D_3},5^+_{\D_4+\D_5-1}) \\
- \f{i}{(4\pi)^2 960}2^5 \f{ \G(\D)}{\left( i \mathcal{D} \right)^\D} \prod_{i=1}^3 (\e_i \s_{i,1})^{\D_i-1} \sum_{k=1}^4 B(\D_4+k-1,\D_5-1) \mathcal{F}^{(1)}_{k}(\{\e_i, z_i,\bar z_i\}) \d\( x - \bar x  \) 
\end{gathered}
\label{order_1_terms_fac1}
\ee
Here we have replaced the $\mathcal{F}^{(1)}_{0}$ dependent term in \eqref{order_1_terms} in terms of a soft graviton mode acting on the 4-point amplitude. Let us now repeat the same procedure for $\mathcal{F}^{(1)}_{1}$. \\
\\
By taking the subleading conformal soft limit $ {\D_4 \to 0} $ in \eqref{order_1_terms_fac1}, we get
\be 
\begin{gathered}
\lim_{\D_4 \to 0} \D_4 \mathcal{M}_5\big|_{\mathcal{O}(1)} = - (\D_5-2) \, \mathcal{H}^1_{-\f{3}{2},\f{1}{2}}(5) \mathcal{M}_4(1^+_{\D_1},2^+_{\D_2},3^+_{\D_3},5^+_{\D_5-1}) \\
- \f{i}{(4\pi)^2 960}2^5 \f{ \G(\sum_{k=1,k\neq 4}^5\D_k)}{\left( i \mathcal{D} \right)^{\sum_{k=1,k\neq 4}^5\D_k}} \prod_{i=1}^3 (\e_i \s_{i,1})^{\D_i-1} \mathcal{F}^{(1)}_{1}(\{\e_i, z_i,\bar z_i\}) \d\( x - \bar x  \) 
\end{gathered}
\label{eq:321}
\ee
Now, from subleading soft graviton theorem we know that:
\be 
\begin{gathered}
\lim_{\D_4 \to 0} \D_4 \mathcal{M}_5 \big|_{\mathcal{O}(1)} = -\mathcal{H}^0_{-1,1}(5) \mathcal{H}^1_{-\f{1}{2},-\f{1}{2}}(5) \mathcal{M}_4(1^+_{\D_1},2^+_{\D_2},3^+_{\D_3},5^+_{\D_5-1})
\end{gathered}
\label{eq:322}
\ee
Again, consistency of the two equations \eqref{eq:321} and \eqref{eq:322} gives us the function $ \mathcal{F}^{(1)}_{1} $ in terms of the leading and subleading soft modes. Substituting this back in \eqref{order_1_terms_fac1} results in
\be 
\begin{gathered}
\mathcal{M}_5\big|_{\mathcal{O}(1)} = \f{\G(\D_4+1)}{\G(\D_4)}B(\D_4-1,\D_5-1) \mathcal{H}^1_{-\f{3}{2},\f{1}{2}}(5) \mathcal{M}_4(1^+_{\D_1},2^+_{\D_2},3^+_{\D_3},5^+_{\D_4+\D_5-1}) \\
+B(\D_4,\D_5-1) \mathcal{H}^0_{-1,1}(5) \( - \mathcal{H}^1_{-\f{1}{2},-\f{1}{2}}(5) \mathcal{M}_4(1^+_{\D_1},2^+_{\D_2},3^+_{\D_3},5^+_{\D_4+\D_5-1}) \)\\
- \f{i}{(4\pi)^2 960}2^5 \f{ \G(\D)}{\left( i \mathcal{D} \right)^\D} \prod_{i=1}^3 (\e_i \s_{i,1})^{\D_i-1} \sum_{k=2}^4 B(\D_4+k-1,\D_5-1) \mathcal{F}^{(1)}_{k}(\{\e_i, z_i,\bar z_i\}) \d\( x - \bar x  \) 
\end{gathered}
\label{4pt_dec_eq}
\ee
We continue this process till all the $ \mathcal{F}^{(1)}_{k} $'s have been replaced by descendant correlation functions of the soft modes. From the above equation \eqref{4pt_dec_eq}, it is clear that to replace all the $ \mathcal{F}^{(1)}_{k} $'s by the descendant correlation functions of the soft modes, we have to go till sub$^4$leading order in the soft limits of $ \D_4 $. We only write the final result here which is given by,
\be 
\begin{gathered}
\mathcal{M}_5(1^+_{\D_1},2^+_{\D_2},3^+_{\D_3},4^+_{\D_4},5^+_{\D_5})\big|_{\mathcal{O}(1)} = \sum_{k=0}^4 \f{1}{(4-k)!}\f{\G(\D_4+4)}{\G(\D_4+k)}B(\D_4+k-1,\D_5-1)  \\
\times \mathcal{H}^{1-k}_{\f{k-3}{2},\f{k+1}{2}}(5)\(\mathcal{H}^1_{-\f{1}{2},-\f{1}{2}}(5)\)^k \mathcal{M}_4(1^+_{\D_1},2^+_{\D_2},3^+_{\D_3},5^+_{\D_4+\D_5-1})
\end{gathered}
\label{fin_OPEO1}
\ee
Now that we have factorized the $\mathcal{O}(1)$ terms in the 5-point Mellin amplitude completely in terms of soft modes acting on the 4-point amplitude, we can easily extract the $\mathcal{O}(1)$ graviton graviton OPE from the above equation. It is given by,
\be\label{00}
\begin{gathered}
G^{+}_{\D_4}(z_4,\bar z_4)G^+_{\D_5}(z_5,\bar z_5)\big|_{\mathcal{O}(1)}  = \sum_{k=0}^4 \f{1}{(4-k)!}\f{\G(\D_4+4)}{\G(\D_4+k)}B(\D_4+k-1,\D_5-1) \\
\times H^{1-k}_{\f{k-3}{2},\f{k+1}{2}}\(H^1_{-\f{1}{2},-\f{1}{2}}\)^k G^+_{\D_4+\D_5-1}(z_5,\bar z_5) 
\end{gathered}
\ee
We can rewrite \eqref{00} using the null states of MHV-sector. From \eqref{phn}, it is clear that all the soft modes $ H^{1-k}_{\f{k-3}{2},\f{k+1}{2}} $ with $ k=1,\cdots,4 $ can be replaced by the MHV null states $ \{\Phi_k, \ k=1,\cdots,4\} $. Thus, \eqref{fin_OPEO1} in terms of the $ \mathcal{O}(1) $ MHV null states \eqref{phn}, becomes:
\be 
\begin{gathered}
G^{+}_{\D_4}(z_4,\bar z_4)G^+_{\D_5}(z_5,\bar z_5) \big|_{\mathcal{O}(1)} = B(\D_4-1,\D_5-1) H^1_{-\f{3}{2},\f{1}{2}} G^+_{\D_4+\D_5-1}(z_5,\bar z_5) \\
+ \sum_{k=1}^4 \f{1}{(4-k)!}\f{\G(\D_4+4)}{\G(\D_4+k)}B(\D_4+k-1,\D_5-1)\Phi_k(\D_4+\D_5)\\
\end{gathered}
\label{f_OPEO1}
\ee
Thus we see that, the $ \mathcal{O}(1) $ terms in the self dual OPE between two positive helicity outgoing gravitons can completely be written in terms of the $ \mathcal{O}(1) $ MHV OPE and the $ \mathcal{O}(1) $ null states of the MHV sector. \\
\\
Now, as discussed in section \ref{revgs}, we can define a new basis for MHV null states instead of $ \Phi_k $'s. This new basis is given by \eqref{nb1}. For our convenience, let us write \eqref{nb1} here again,
\be
\begin{gathered}
\Omega_k(\D) = \sum_{n=1}^{k} \f{1}{(k -n)!} \f{\G(\D+k-2)}{\G(\D +n-2)}\Phi_n(\D)
\end{gathered} 
\label{nbm1}
\ee
This basis has nice transformation properties under the $ w $-algebra \cite{Banerjee:2023zip}, reviewed in section \ref{mnt}. Represented in terms of this new $ \Omega $-basis, the graviton-graviton OPE \eqref{f_OPEO1} takes a very simple form,
\be 
\begin{gathered}
G^{+}_{\D_4}(z_4,\bar z_4)G^+_{\D_5}(z_5,\bar z_5) \big|_{\mathcal{O}(1)} = B(\D_4-1,\D_5-1) H^1_{-\f{3}{2},\f{1}{2}} G^+_{\D_4+\D_5-1}(z_5,\bar z_5) \\
+ \sum_{k=1}^4 B(\D_4+k-1,\D_5-1)\Omega_k(\D_4+\D_5)\\
\end{gathered}
\label{fm_OPEO1}
\ee
Thus we see that, the $ \mathcal{O}(1) $ OPE between two positive helicity outgoing gravitons in quantum self dual gravity truncates at $ n=4 $ of the general $ w $-invariant OPE \eqref{0}. We now move on to $  \mathcal{O} (\bar z) $ OPE in the next subsection.

\subsubsection{$  \mathcal{O} (\bar z_{45})$ Term}

The soft modes that appear at order $ \bar z_{45} $ from the $  w $-algebra are given by 
\be
H^{k}_{-\f{k+2}{2},-\f{k}{2}}, \ k=1,0,-1,\cdots 
\ee
Now, like the $ \mathcal{O}(1) $ OPE we can factorize the $  \mathcal{O} (\bar z_{45})$ terms from the 5-point amplitude using the soft limits and $ w $-modes. The crucial difference from $ \mathcal{O}(1) $ is that, now we have to go one order higher in the soft limits than $ \mathcal{O}(1) $. We start by writing the $ \mathcal{O}(\bar z_{45}) $ term of the 5-point Mellin amplitude given by (see \eqref{order_bz_app_2}),
\be
\begin{gathered}
\mathcal{M}_5\big|_{\mathcal{O}(\bar z_{45})} = - \f{i}{(4\pi)^2 960}2^5 \f{\G(\D)}{\(i\mathcal{D}\)^\D} \prod_{i=1}^3 (\e_i \s_{i,1})^{\D_i-1} \sum_{k=1}^5 B(\D_4+k-1,\D_5-1) \mathcal{F}^{(\bar z)}_k (\{\e_i, z_i,\bar z_i\})
\end{gathered} 
\label{order_bz_main}
\ee
One can easily see from \eqref{order_bz_main} that, to factorize the 5-point Mellin amplitude completely, i.e, to replace all the functions $ \mathcal{F}^{(\bar z)}_k (\{\e_i, z_i,\bar z_i\}) $ by the descendant correlation functions of soft modes, we have to continue taking the soft limits in $ \D_4 $ till we reach $ \D_4 \to -4 $. Thus the highest soft modes that can appear in the OPE at $ \mathcal{O}(\bar z_{45}) $ are given by $ H^{-4}_{1,2} $. We have discussed how to factorize the amplitude at $ \mathcal{O}(1) $ in terms of the descendant correlators of the soft modes in the previous section in detail. One has to repeat the same procedure for $ \mathcal{O}(\bar z_{45}) $ as well. Without going into much detail we directly write the $ \mathcal{O}(\bar z_{45}) $ OPE which is given by,
\be
\begin{gathered}
G^+_{\D_4}(z_4, \bar z_4)G^+_{\D_5}(z_5,\bar z_5)\big|_{\mathcal{O}(\bar z_{45})} = G^{+}_{\D_4}(z_4,\bar z_4)G^+_{\D_5}(z_5, \bar z_5)\big|_{\text{MHV at} \  \mathcal{O}(\bar z_{45})} \\
+ \sum_{k=1}^4 \f{1}{(n-k)!}\f{\G(\D_4+n+1)}{\G(\D_4+k+1)}B(\D_4+k,\D_5-1) \Psi_k(\D_4+\D_5+1)
\label{ozbOPE}
\end{gathered} 
\ee
where
\be \label{mhvzb}
\begin{gathered}
G^{+}_{\D_4}(z_4,\bar z_4)G^+_{\D_5}(z_5, \bar z_5)\big|_{\text{MHV at} \  \mathcal{O}(\bar z_{45})} = B(\D_4-1,\D_5-1)\[\f{\D_4-1}{\D_4+\D_5-2}H^{0}_{-1,0}\(-H^{1}_{-\f{1}{2},-\f{1}{2}}\) \right.\\
\left. + \D_4 \, H^{1}_{-\f{3}{2},-\f{1}{2}}\] G^+_{\D_4+\D_5-1}
\end{gathered}
\ee
and $\Psi_k(\D_4+\D_5+1)$ is given by
\bea
\no \Psi_k(\D_4+\D_5+1) &=& \[H^{-k}_{\f{k-2}{2},\f{k}{2}}\(-H^{1}_{-\f{1}{2},-\f{1}{2}}\)^{k+1} - \f{(-1)^k}{k!}\f{\G(\D_4+\D_5+k-1)}{\G(\D_4+\D_5-1)} H^{0}_{-1,0}\(-H^{1}_{-\f{1}{2},-\f{1}{2}}\) \right.\\
&& \left. - (-1)^k \f{k}{(k+1)!} \f{\G(\D_4+\D_5+k-1)}{\G(\D_4+\D_5-2)} H^{1}_{-\f{3}{2},-\f{1}{2}}\] G^+_{\D_4+\D_5-1}
\eea
In terms of the new basis defined in \eqref{nb2}, the above OPE can again be written in a very nice and simple form given by,
\be\label{ozm}
\begin{gathered}
G^{+}_{\D_4}(z_4,\bar z_4)G^+_{\D_5}(z_5,\bar z_5)\big|_{\mathcal{O}(\bar z_{45})} =  G^{+}_{\D_4}(z_4,\bar z_4)G^+_{\D_5}(z_5, \bar z_5)\big|_{\text{MHV at} \  \mathcal{O}(\bar z_{45})} \\
+ \bar z_{45} \sum_{k=1}^4 B(\D_4+k,\D_5-1) \ \Pi_k(\D_4+\D_5+1)
\end{gathered} 
\ee
Thus we see that the $ \mathcal{O}(\bar z_{45}) $ terms in the OPE again truncate at $ n=4 $ of the general $ w $-invariant OPE \eqref{0}.

\section{Discussion}
Operator product expansion plays a very important role in any quantum field theory and therefore it is important to understand the structure of OPE in the celestial CFTs. In its current formulation, celestial CFTs differ from more conventional CFTs in many ways. The primary difference is that the spectrum of the operator dimensions in celestial CFTs is not bounded from below. Taken at face value, this implies that the number of descendants that can appear at any given order of the celestial OPE can be \textit{infinite}. However, this is not a very desirable feature and warrants further study. 

In this paper, we have undertaken the task of computing the celestial OPE of two positive helicity outgoing gravitons in the quantum self-dual gravity. It is known that the self dual gravity enjoys $w$ invariance. Therefore, one should be able to express the OPE in terms of $w$ descendants of the graviton primary. This is what we have found. However, the most surprising fact which comes out of our study is that at any given order the OPE contains only a \textit{finite} number of $w$ descendants. Therefore, the self dual gravity behaves like any other CFT with a spectrum of operator dimensions bounded from below. 

This raises some interesting questions. For example, we know that the Holographic Symmetry Algebra (HSA) contains an infinite tower of holomorphic currents $H^k_n(z)$ with $k$ going from $1$ to $-\infty$. Our calculation shows that in the self dual theory at $\mathcal{O}(1)$ and at $\mathcal{O}(\bar z)$ the list of $w$ descendants truncate at $k=-3$ and $k=-4$, respectively. However, this is somewhat unnatural given the fact that the currents $H^{-3}_n(z)$ and $H^{-4}_n(z)$ do not play any distinguished role in the algebra. Therefore, it is natural to wonder if there are other $w$ invariant theories where the truncation occurs at other values of $k$. This is consistent with our earlier analysis \cite{Banerjee:2023zip} based on the $w$ algebra representation where we found that one can write down an infinite number of consistent $w$ invariant OPEs where truncation happens at different values of $k$. Therefore, truncation is not a reflection of $w$ symmetry. We leave the construction of these theories as an interesting problem for the future. 

Before we end, we would like to point out that truncation means that the self dual theory in many ways behave like theories with operator dimensions bounded from below. So it is very likely that the self dual theory and the (tree-level) MHV sector of GR can be reformulated in terms of celestial primary operators with dimensions strictly bounded from below. Interesting proposals along this line has been put forward in \cite{Cotler:2023qwh,Freidel:2022skz} It will be fascinating if they can be applied to the present problem. 

\section{Acknowledgements}

SB would like to thank the participants of the Kickoff Workshop for the Simons Collaboration on Celestial Holography for helpful comments. The work of SB is partially supported by the  Swarnajayanti Fellowship (File No- SB/SJF/2021-22/14) of the Department of Science and Technology and SERB, India. The work of HK is partially supported by the KVPY fellowship of the Department of Science and Technology, Government of India. The work of PP is supported by an IOE endowed Postdoctoral position at IISc, Bengaluru, India.

\appendix

\section{Brief Review of Celestial or Mellin Amplitudes For Massless Particles}\label{review}

The Celestial or Mellin amplitude for massless particles in four dimensions is defined as the Mellin transformation of the $S$-matrix element, given by \cite{Pasterski:2016qvg,Pasterski:2017kqt}
\be\label{mellin}
\mathcal M_n\big(\{z_i, \bar z_i, h_i, \bar h_i\}\big) = \prod_{i=1}^{n} \int_{0}^{\infty} d\omega_i \ \omega_i^{\D_i -1} A_n\big(\{\omega_i,z_i,\bar z_i, \sigma_i\}\big)
\ee 
where $\sigma_i$ denotes the helicity of the $i$-th particle and the on-shell momenta are parametrized by \eqref{mom}. The scaling dimensions $(h_i,\bar h_i)$ are defined as,

\be
h_i = \frac{\D_i + \sigma_i}{2}, \quad \bar h_i = \frac{\D_i - \sigma_i}{2}
\ee
Under the Lorentz transformation \eqref{lt}, the Mellin amplitude $\mathcal M_n$ transforms as,
\be
\mathcal M_n\big(\{z_i, \bar z_i, h_i, \bar h_i\}\big) = \prod_{i=1}^{n} \frac{1}{(cz_i + d)^{2h_i}} \frac{1}{(\bar c \bar z_i + \bar d)^{2\bar h_i}} \mathcal M_n\bigg(\frac{az_i+b}{cz_i+d} \ ,\frac{\bar a \bar z_i + \bar b}{\bar c \bar z_i + \bar d} \ , h_i,\bar h_i\bigg)
\ee
This is the familiar transformation law for the correlation function of primary operators of weight $(h_i,\bar h_i)$ in a $2$-D CFT under the global conformal group.

In Einstein gravity, the Mellin amplitude as defined in \eqref{mellin} usually diverges. This divergence can be regulated by defining a modified Mellin amplitude as \cite{Banerjee:2018gce,Banerjee:2019prz}, 
\be\label{mellinmod}
\mathcal M_n\big(\{u_i,z_i, \bar z_i, h_i, \bar h_i\}\big) = \prod_{i=1}^{n} \int_{0}^{\infty} d\omega_i \ \omega_i^{\D_i -1} e^{-i\sum_{i=1}^n \epsilon_i \omega_i u_i} A_n\big(\{\omega_i,z_i,\bar z_i, \sigma_i\}\big)
\ee 
where $u$ can be thought of as a time coordinate and $\epsilon_i = \pm 1$ for an outgoing (incoming) particle. Under (Lorentz) conformal tranansformation the modified Mellin amplitude $\mathcal M_n$ transforms as,
\be
\mathcal M_n\big(\{u_i,z_i, \bar z_i, h_i, \bar h_i\}\big) = \prod_{i=1}^{n} \frac{1}{(cz_i + d)^{2h_i}} \frac{1}{(\bar c \bar z_i + \bar d)^{2\bar h_i}} \mathcal M_n\bigg(\frac{u_i}{|cz_i + d|^2} \ , \frac{az_i+b}{cz_i+d} \ ,\frac{\bar a \bar z_i + \bar b}{\bar c \bar z_i + \bar d} \ , h_i,\bar h_i\bigg)
\ee
Under global space-time translation, $u \rightarrow u + A + Bz + \bar B\bar z + C z\bar z$, the modified Mellin amplitude is invariant, i.e, 
\be
\mathcal M_n\big(\{u_i + A + Bz_i + \bar B\bar z_i + C z_i\bar z_i ,z_i, \bar z_i, h_i, \bar h_i\}\big) = \mathcal M_n\big(\{u_i,z_i, \bar z_i, h_i, \bar h_i\}\big)
\ee

Now in order to make manifest the conformal nature of the dual theory living on the celestial sphere it is useful to write the (modified) Mellin amplitude as a correlation function of conformal primary operators. So let us define a generic conformal primary operator as, 
\be
\label{confprim}
\phi^{\epsilon}_{h,\bar h}(z,\bar z) = \int_{0}^{\infty} d\omega \  \omega^{\D-1} a(\epsilon\omega, z, \bar z, \sigma)
\ee
where $\epsilon=\pm 1$ for an annihilation (creation) operator of a massless particle of helicity $\sigma$. Under (Lorentz) conformal transformation the conformal primary transforms like a primary operator of scaling dimension $(h,\bar h)$
\be
\phi'^{\epsilon}_{h,\bar h}(z,\bar z) = \frac{1}{(cz + d)^{2h}} \frac{1}{(\bar c \bar z + \bar d)^{2\bar h}} \mathcal \phi^{\epsilon}_{h,\bar h}\bigg(\frac{az+b}{cz+d} \ ,\frac{\bar a \bar z + \bar b}{\bar c \bar z + \bar d}\bigg)
\ee
Similarly in the presence of the time coordinate $u$ we have,
\be
\label{confprimu}
\phi^{\epsilon}_{h,\bar h}(u,z,\bar z) = \int_{0}^{\infty} d\omega \ \omega^{\D-1} e^{-i \epsilon \omega u} a(\epsilon\omega, z, \bar z, \sigma)
\ee
Under (Lorentz) conformal transformations 
\be
\phi'^{\epsilon}_{h,\bar h}(u,z,\bar z) = \frac{1}{(cz + d)^{2h}} \frac{1}{(\bar c \bar z + \bar d)^{2\bar h}} \mathcal \phi^{\epsilon}_{h,\bar h}\bigg(\frac{u}{|cz+d|^2},\frac{az+b}{cz+d} \ ,\frac{\bar a \bar z + \bar b}{\bar c \bar z + \bar d}\bigg)
\ee


In terms of \eqref{confprim},  the Mellin amplitude can be written as the correlation function of conformal primary operators
\be
\mathcal M_n = \bigg\langle{\prod_{i=1}^n \phi^{\epsilon_i}_{h_i,\bar h_i}(z_i,\bar z_i)}\bigg\rangle
\ee
Similarly using \eqref{confprimu}, the modified Mellin amplitude can be written as,
\be
\mathcal M_n = \bigg\langle{\prod_{i=1}^n \phi^{\epsilon_i}_{h_i,\bar h_i}(u_i,z_i,\bar z_i)}\bigg\rangle
\ee 

\subsection{Comments on notation in the paper}
Note that the conformal primaries carry an extra index $\epsilon$ which distinguishes between an incoming and an outgoing particle. In this paper, for notational simplicity, we omit this additional index unless this plays an important role. So in most places we simply write the (modified) Mellin amplitude as,
\be
\mathcal M_n = \bigg\langle{\prod_{i=1}^n \phi_{h_i,\bar h_i}(z_i,\bar z_i)}\bigg\rangle
\ee
or
\be
\mathcal M_n = \bigg\langle{\prod_{i=1}^n \phi_{h_i,\bar h_i}(u_i,z_i,\bar z_i)}\bigg\rangle
\ee 
Similarly in many places in the paper we denote a graviton primary of weight $\D = h+\bar h$ by $G^{\sigma}_\D$ where $\sigma = \pm 2$ is the helicity (= $h-\bar h$). Since we are considering pure gravity, we can further simplify the notation to $G^{\pm}_{\D}$ by omitting the $2$. 

\section{Parametrization of the Delta Functions}
\label{dp}
In this Appendix, we parametrize the 4-point and 5-point delta functions which will be convenient for our purpose of extracting the OPE. 

\subsection{4-Point Delta Function}

In $ (2,2) $ split signature, the parametrization of the null momentum ($ p_i $) for $ i $-th massless particle in terms of $ \(\om_i, z_i, \bar z_i \) $ is given by
\be 
p_i = \om_i \{ 1+z_i \bar{z}_i, z_i+\bar{z}_i, (z_i-\bar{z}_i), 1-z_i \bar{z}_i\}, \qquad p_i^2 = 0
\label{mom_param}
\ee
This allows us to write down the 4-point momentum conserving delta function in the following way which is more convenient for us
\bea
\no \d^{(4)}\(\sum_{i=1, \neq 4}^5 \e_i p_i \) &=& \f{1}{4}\d\(\sum_{i=1, \neq 4}^5 \e_i \om_i \) \d\(\sum_{i=1}^3 \e_i \om_i z_{i5}\) \d\(\sum_{i=1}^3 \e_i \om_i \bar{z}_{i5}\) \d\(\sum_{i=1}^3 \e_i \om_i z_{i5} \bar{z}_{i5}\)\\
\no &=& \e_1 \e_2 \e_3 \e_5 \f{1}{4 \om_5}\d(\om_1 - \om_1^*)\d(\om_2 - \om_2^*)\d(\om_3 - \om_3^*)\d(x - \bar x)\\
\label{4pt_delta}
\eea
where $ \e_i = \pm 1 $ for outgoing (incoming) particle and 
\bea
\om_i^* &=&  \e_5 \, \om_5 \, \e_i \, \s_{i,1}\\
\label{sig11}
\s_{1,1} &=& -\f{ z_{25} \bar z_{35}}{z_{12}\bar z_{13}}\\
\label{sig21}
\s_{2,1} &=& \f{ z_{15} \bar z_{35}}{z_{12}\bar z_{23}}\\
\label{sigma31}
\s_{3,1} &=& -\f{z_{25} \bar z_{15}}{z_{23}\bar z_{13}}\\
\label{sig41}
x &=& z_{12}z_{35} \bar z_{13} \bar z_{25}, \ \bar x = z_{13}z_{25} \bar z_{12} \bar z_{35}
\label{xxb}
\eea
The $ \s_{i,1} $'s defined above satisfy the following identities on the support of $ \d(x - \bar x) $
\bea
\s_{1,1} + \s_{2,1} + \s_{3,1} + 1 &=& 0\\
z_{15} \s_{1,1} + z_{25} \s_{2,1} + z_{35}\s_{3,1} &=& 0\\
\bar z_{15} \s_{1,1} + \bar z_{25} \s_{2,1} + \bar z_{35}\s_{3,1} &=& 0
\eea
This representation for the 4-point delta function and the properties of $\s_{i,1}$'s will be useful in extracting the OPE. Note that in this delta function representation, we have indexed the four particles by $1,2,3,5$ because to extract the OPE, we take the 4 $\to$ 5 OPE limit in the 5-point Mellin amplitude and then factorize it in terms of the 4-point Mellin amplitude now indexed by $1,2,3,5$. This is a notation that we followed throughout the paper.



\subsection{5-Point Delta Function}
\label{fpd}

We now write down the representation for the delta function for 5 particles. For concreteness, we take $ \e_4=\e_5=+1 $. Since we are interested in the OPE limit $ 4\to 5 $, it is convenient to use the following parametrization
\be
\om_4 = t\om_P , \ \om_5 = (1-t)\om_P 
\ee
in representing the 5-point delta function. For the case of $n=5$ particles in four spacetime dimensions we have four constraint equations coming from the four components of the energy momentum conserving equations. We can solve these four constraint equations for three energy variables $ \{\om_1,\om_2,\om_3\} $ in terms of $ \om_4 $ and $ \om_5 $. Thus, the representation of the 5-point delta function which is better suited for our purposes of performing the OPE decomposition of the Mellin amplitude in the (4, 5) channel, is given by \cite{Banerjee:2020zlg} \footnote{Please note that in \cite{Banerjee:2020zlg} the OPE factorization has been done starting from the 6-point Mellin amplitude whereas in this paper it is done starting from the 5-point amplitude. Thus, in parametrizing the 5-point delta function in this paper, we have used the same methodology which was used for 6-point delta function in \cite{Banerjee:2020zlg}.},
\bea
\no \d^{(4)}\(\sum_{i=1}^5 \e_i \om_i q_i\) &=& \f{1}{4 \om_P}\d(\om_1-\om_1^*)\d(\om_2-\om_2^*)\d(\om_3-\om_3^*)\\
\no && \times \d\( x - \bar x - t z_{45}\(\f{x}{z_{35}}-\f{\bar x}{z_{25}}\) - t \bar z_{45}\(\f{x}{\bar z_{25}}-\f{\bar x}{\bar z_{35}}\) + t z_{45} \bar z_{45}\(\f{x}{z_{35}\bar z_{25}}-\f{\bar x}{z_{25}\bar z_{35}}\) \) \\
 \label{5pt_delta_fn}
\eea
where for $ i=\{1,2,3\} $ we have
\bea
\no \om_i^* &=& \om_P \tilde{\om}_i^*\\
\tilde{\om}_i^* &=& \e_i \(\s_{i,1} + t z_{45} \s_{i,2} + t \bar z_{45}\s_{i,3} + t z_{45}\bar z_{45} \s_{i,4}\)
\label{om_st}
\eea
and the $ \s_{i,1}, \ x, \ \bar x $ are given by \eqref{sig11}-\eqref{xxb}. We also have 
\be
\s_{i,2} = \f{\pa \s_{i,1}}{\pa z_5}, \ \s_{i,3} = \f{\pa \s_{i,1}}{\pa \bar z_5}, \ \s_{i,4} = \f{\pa \s_{i,1}}{\pa z_5 \pa \bar z_5}, \qquad \forall \, i=1,2,3.
\ee

\section{Simplification of the 4-point Amplitude}
\label{4pt_sd_mom_simp}

In this Appendix, we simplify the 4-point self dual one loop amplitude in momentum space which is used in section \ref{OPEA}. We start with the equation \eqref{frpt} for the 4-point amplitude:
\bea
\no B_4 &=&  \sum_{\substack{1\leq a <b \leq 4 \\ M,N}} h(a,M,b) h(b,N,a) \text{\text{tr}}^3[aMbN]\\
\no &=& h(1,3,2) h(2,4,1) \text{tr}^3[1324] + h(1,2,3) h(3,4,1) \text{tr}^3[1234] + h(1,2,4) h(4,3,1) \text{tr}^3[1243]\\
\no &+& h(2,1,3) h(3,4,2) \text{tr}^3[2134] + h(2,1,4) h(4,3,2) \text{tr}^3[2143] + h(3,1,4) h(4,2,3) \text{tr}^3[3142] 
\label{B_41}
\eea
The trace function is given by
\be
\text{tr}[aMbN] = \left< a |K_M|b \right]\left<b|K_N|a\right]+ \left[ a |K_M|b \right> \left[b|K_N|a\right> 
\ee
For $M=\{i\},\ N=\{l\}$ we have
\bea
\no \text{tr}[aibl] &=& \left< a |k_i|b \right]\left<b|k_l|a\right]+ \left[ a |k_i|b \right> \left[b|k_l|a\right>\\
&=& \left< ai \right> [ib]\left< bl \right> [la] + \left< bi \right> [ia] \left< al \right> [lb]
\label{trace_fn}
\eea
From the above equation we can see that $ \text{tr}[aibl] = \text{tr}[ialb] $. Using this property of the trace function and the expression for the half soft function 
\be
h(a,i,b) = \f{1}{\left< ai \right>^2 \left< ib \right>^2 }
\label{Bt_4}
\ee
\eqref{B_41} can be simplified as,
\bea
\no B_4 &=& 2\( h(1,3,2) h(2,4,1) \text{tr}^3[1324] + h(1,2,3) h(3,4,1) \text{tr}^3[1234] + h(1,2,4) h(4,3,1) \text{tr}^3[1243] \)\\
\label{B_42}
\eea
Now, using the momentum conservation for four particles in the trace functions \eqref{trace_fn} and the explicit expressions of the half soft functions \eqref{Bt_4}, \eqref{B_42} finally gives,
\bea
\no B_4 &=& -2^4\( \f{\<13\>\<23\>([13][23])^3}{\<14\>^2\<24\>^2} + (1 \leftrightarrow 3)  + (2 \leftrightarrow 3) \)\\
\label{B_43}
\eea
As mentioned earlier, since we will index the 4 particles as $1,2,3,5$, relabelling $ 4 $ as $ 5 $ in the above expression gives the following form of the 4-point amplitude in momentum space, 
\bea
\no B_4 &=& -2^4\( \f{\<13\>\<23\>([13][23])^3}{\<15\>^2\<25\>^2} + (1 \leftrightarrow 3)  + (2 \leftrightarrow 3) \)\\
\label{B_44}
\eea

\section{Simplification of the 5-point Amplitude}
\label{5pt_sd_mom_simp}

Similar to what was done for the 4-point case, we will now simplify the 5-point self dual one loop amplitude in momentum space which is used in section \ref{5pt_sd_mom} by considering the equation \eqref{5pt_sdg_ly_b5}:
\bea
\no B_5 &=&  \sum_{\substack{1\leq a <b \leq 5 \\ M,N}} h(a,M,b) h(b,N,a) \text{tr}^3[aMbN]\\
\no &=& h(1,M,2) h(2,N,1) \text{tr}^3[1M2N] + h(1,M,3) h(3,N,1) \text{tr}^3[1M3N] + h(1,M,4) h(4,N,1) \text{tr}^3[1M4N]\\
\no &+& h(1,M,5) h(5,N,1) \text{tr}^3[1M5N] + h(2,M,3) h(3,N,2) \text{tr}^3[2M3N] + h(2,M,4) h(4,N,2) \text{tr}^3[2M4N]\\
\no &+& h(2,M,5) h(5,N,2) \text{tr}^3[2M5N] + h(3,M,4) h(4,N,3) \text{tr}^3[3M4N] + h(3,M,5) h(5,N,3) \text{tr}^3[3M5N]\\
&+& h(4,M,5) h(5,N,4) \text{tr}^3[4M5N]
\label{B_51}
\eea
The two sets $M$ and $N$ are such that $M \bigcup N = {1,...a-1,a+1,...b-1,b+1,...n}$ and $M \bigcap N = \phi$ and the sum is over all possible $a,b$ and sets $(M,N)$, where $(M,N)$ and $(N,M)$ are not distinguished.
For 5-point amplitudes, with $M=\{i,j\},\ N=\{l\}$, the trace function given by,
\be
\text{tr}[aMbN] = \left< a |K_M|b \right]\left<b|K_N|a\right]+ \left[ a |K_M|b \right> \left[b|K_N|a\right> 
\ee
becomes
\bea
\no \text{tr}[a\{i+j\}b\{l\}] &=& \left< a |k_i+k_j|b \right]\left<b|k_l|a\right]+ \left[ a |k_i+k_j|b \right> \left[b|k_l|a\right>\\
\no &=& \text{tr}[aibl] + \text{tr}[ajbl]\\
&=& (\left< ai \right> [ib]  + \left< aj \right> [jb])\left< bl \right> [la] + (\left< bi \right> [ia] + \left< bj \right> [ja])\left< al \right> [lb]
\eea
Now, using momentum conservation in the spinor notation
\bea
\no && \left< ai(\neq \{a,b,j,k\}) \right> [i(\neq \{a,b,j,k\}) b] + \left< aj(\neq \{a,b,i,k\})  \right> [j(\neq \{a,b,j,k\}) b] \\
 && + \left< ak(\neq \{a,b,i,j\})  \right> [k(\neq \{a,b,i,j\}) b] = 0
\eea
one can show that,
\bea
\no \text{tr}[a\{i+j\}b\{l\}] &=& - 2\left< al \right>[al]\left< bl \right> [bl]
\eea
where each label is different. Thus we see that $ \text{tr}[a\{i+j\}b\{l\}] $ is independent of $ \{i,j\} $.
The half soft functions needed for the simplification of the 5-point amplitude are given by,
\bea
h(a,\{i,j\},b) &=& \f{[ij]}{\left<ij\right> \left< ai \right> \left< aj \right> \left< ib \right>\left< jb \right>}\\
h(a,\{i\},b) &=& \f{1}{\left< ai \right>^2 \left< ib \right>^2 }
\eea
Thus, we see that $ h(a,\{i,j\},b) = h(a,\{j,i\},b) $. \\
\\
Now, using the explicit form of the trace and half soft functions in terms of spinor helicity brackets, we can write \eqref{B_51} as, 
\bea
\no B_5 = -8 \left[ \f{[25]\left< 13 \right>\left< 34 \right> ([13][34])^3}{\lt<25\rt>\left< 12 \right> \left< 15 \right> \left< 24 \right>\left< 54 \right>} + \f{[24]\left< 13 \right>\left< 35 \right>([13] [35])^3}{\lt<24\rt>\left< 12 \right> \left< 14 \right> \left< 25 \right>\left< 45 \right>} + \f{[15]\left< 23 \right>\left< 34 \right>([23] [34])^3}{\lt<15\rt>\left< 21 \right> \left< 25 \right> \left< 14 \right>\left< 54 \right>} \right. \\
\no + \left.\f{[14]\left< 23 \right>\left< 35 \right>([23] [35])^3}{\lt<14\rt>\left< 21 \right> \left< 24 \right> \left< 15 \right>\left< 45 \right>} + \f{[45]\left< 13 \right>\left< 23 \right>([13] [23])^3}{\lt<45\rt>\left< 14 \right> \left< 15 \right> \left< 42 \right>\left< 52 \right>} + \f{[34]\left< 25 \right>\left< 15 \right>([15] [25])^3}{\lt<34\rt>\left< 13 \right> \left< 14 \right> \left< 32 \right>\left< 42 \right>} \right.\\
\no + \left. \f{[35]\left< 14 \right>\left< 24 \right>([14] [24])^3}{\lt<35\rt>\left< 13 \right> \left< 15 \right> \left< 32 \right>\left< 52 \right>} + \f{[12]\left< 34 \right>\left< 35 \right>([34] [35])^3}{\lt<12\rt>\left< 41 \right> \left< 42 \right> \left< 15 \right>\left< 25 \right>} + \f{[12]\left< 35 \right>\left< 45 \right>([35] [45])^3}{\lt<12\rt>\left< 31 \right> \left< 32 \right> \left< 14 \right>\left< 24 \right>} \right. \\
\no \left. + \f{[12]\left< 34 \right>\left< 45 \right>([34] [45])^3}{\lt<12\rt>\left< 31 \right> \left< 32 \right> \left< 15 \right>\left< 25 \right>} \right] + \( 1 \leftrightarrow 3 \) + \( 2 \leftrightarrow 3 \) \\
\label{B5_30terms}
\eea
Before simplifying this, first note that the first 4 terms (and hence a total of 12 terms) in the above expression have the apparent form that seems to go like $\sim \frac{1}{\langle 45 \rangle}$. However, it cannot be true that the 5-point amplitude has a leading behaviour of $\sim \frac{1}{\langle 45 \rangle}$. We will show that these terms add up to contribute to the leading order ($\mathcal{O}(\frac{[45]}{\langle 45 \rangle})$), $\mathcal{O}(1)$ and higher orders as expected. Hence, to simplify further, let us first write down these 12 terms explicitly,
\bea
\no -\frac{B_5^{\tn{S}}}{8} &=& \f{[25]\left< 13 \right>\left< 34 \right> ([13][34])^3}{\lt<25\rt>\left< 12 \right> \left< 15 \right> \left< 24 \right>\left< 54 \right>} + \f{[35]\left< 12 \right> \left< 24 \right> ([12][24])^3 }{\lt<35\rt>\left< 13 \right> \left< 15 \right> \left< 34 \right>\left< 54 \right>} + \f{[24]\left< 13 \right>\left< 35 \right>([13] [35])^3}{\lt<24\rt>\left< 12 \right> \left< 14 \right> \left< 25 \right>\left< 45 \right>} \\
\no &+& \f{[34]\left< 12 \right>\left< 25 \right>([12] [25])^3}{\lt<34\rt>\left< 13 \right> \left< 14 \right> \left< 35 \right>\left< 45 \right>} + \f{[15]\left< 23 \right>\left< 34 \right>([23] [34])^3}{\lt<15\rt>\left< 21 \right> \left< 25 \right> \left< 14 \right>\left< 54 \right>} + \f{[35]\left< 12 \right>\left< 14 \right>([12] [14])^3}{\lt<35\rt>\left< 23 \right> \left< 25 \right> \left< 34 \right>\left< 54 \right>} \\
\no &+& \f{[14] \left< 23 \right>\left< 35 \right>([23] [35])^3}{\lt<14\rt>\left< 21 \right> \left< 24 \right> \left< 15 \right>\left< 45 \right>} + \f{[34]\left< 12 \right>\left< 15 \right>([12] [15])^3}{\lt<34\rt>\left< 23 \right> \left< 24 \right> \left< 35 \right>\left< 45 \right>} + \f{[15]\left< 23 \right>\left< 24 \right>([23] [24])^3}{\lt<15\rt>\left< 31 \right> \left< 35 \right> \left< 14 \right>\left< 54 \right>} \\
\no &+& \f{[25]\left< 13 \right>\left< 14 \right> ([13][14])^3}{\lt<25\rt>\left< 32 \right> \left< 35 \right> \left< 24 \right>\left< 54 \right>} + \f{[14]\left< 23 \right>\left< 25 \right>([23] [25])^3}{\lt<14\rt>\left< 31 \right> \left< 34 \right> \left< 15 \right>\left< 45 \right>} + \f{[24]\left< 13 \right>\left< 15 \right>([13] [15])^3}{\lt<24\rt>\left< 32 \right> \left< 34 \right> \left< 25 \right>\left< 45 \right>} \\
\label{sing_terms}
\eea
Keeping terms only upto $\mathcal{O}(\bar{z}_{45})$, the first term above can be rewritten as,
\begin{equation}
\begin{gathered}
\f{[25]\left< 13 \right>\left< 34 \right> ([13][34])^3}{\lt<25\rt>\left< 12 \right> \left< 15 \right> \left< 24 \right>\left< 54 \right>} = - \f{1}{\<45\>}\f{[25]\<12\>^3\<24\>^3[12]^3[24]^3}{\<25\> \<12\> \<15\> \<24\>\<13\>^2\<34\>^2} + 3\f{[45]}{\<45\>}\f{[25]\<12\>^2\<15\>\<24\>^3[12]^3[24]^2}{\<25\> \<12\> \<15\> \<24\>\<13\>^2\<34\>^2} \\
+ 3\f{[25]\<12\>^3\<24\>^2[12]^2 [15] [24]^3}{\<25\> \<12\> \<15\> \<24\>\<13\>^2\<34\>^2} - 3\left<45\right>\f{[25]\<12\>^2[12] [15]^2 [24]^3}{\<25\> \<15\> \<13\>^2\<34\>^2} - 9\[45\]\f{[25]\<12\> \<24\> [12]^2 [15] [24]^2}{\<25\> \<13\>^2\<34\>^2}
\label{first_sing_term}
\end{gathered}
\end{equation}
Now we use a little trick to explicitly show that the terms in \eqref{sing_terms} add up to give ($\mathcal{O}(\frac{[45]}{\langle 45 \rangle})$), $\mathcal{O}(1)$ and higher orders contributions. It involves appropriately combining terms in the equation. To see this, note that the first term in RHS of \eqref{first_sing_term} and second term in RHS of \eqref{sing_terms} can be combined to get,
\begin{equation}
\begin{gathered}
- \f{1}{\<45\>}\f{[25]\<12\>^3\<24\>^3[12]^3[24]^3}{\<25\> \<12\> \<15\> \<24\>\<13\>^2\<34\>^2} + \f{[35]\left< 12 \right> \left< 24 \right> ([12][24])^3 }{\lt<35\rt>\left< 13 \right> \left< 15 \right> \left< 34 \right>\left< 54 \right>}\\
= - \f{1}{\<45\>}\f{\left< 12 \right> \left< 24 \right> ([12][24])^3 }{\left< 13 \right>^2 \left< 15 \right>\<25\> \left< 34 \right>^2 \left< 35 \right>}\(\<12\>\<24\>\<35\>[25]+\<13\>\<34\>\<25\>[35]\)
\end{gathered}
\end{equation}
Note that although we are writing equalities everywhere, one should keep in mind that there are higher order terms as well. However, here, and throughout this paper, we will always write expressions keeping terms only upto $\mathcal{O}(\bar{z}_{45})$. Now, using the Shouten Identity $ \<24\>\<35\> = \<25\>\<34\> + \<23\>\<45\> $ and momentum conservation equation, we can write the above equation as
\be 
\begin{gathered}
- \f{1}{\<45\>}\f{[25]\<12\>^3\<24\>^3[12]^3[24]^3}{\<25\> \<12\> \<15\> \<24\>\<13\>^2\<34\>^2} + \f{[35]\left< 12 \right> \left< 24 \right> ([12][24])^3 }{\lt<35\rt>\left< 13 \right> \left< 15 \right> \left< 34 \right>\left< 54 \right>}\\
= - \f{1}{\<45\>}\f{\left< 12 \right> \left< 24 \right> ([12][24])^3 }{\left< 13 \right>^2 \left< 15 \right>\<25\> \left< 34 \right>^2 \left< 35 \right>}\(-\<14\>\<25\>\<34\>[45]+\<12\>\<23\>\<45\>[25]\)
\end{gathered}
\ee
Hence the first two terms in \eqref{sing_terms} give
\be
\begin{gathered}
\f{[25]\left< 13 \right>\left< 34 \right> ([13][34])^3}{\lt<25\rt>\left< 12 \right> \left< 15 \right> \left< 24 \right>\left< 54 \right>} + \f{[35]\left< 12 \right> \left< 24 \right> ([12][24])^3 }{\lt<35\rt>\left< 13 \right> \left< 15 \right> \left< 34 \right>\left< 54 \right>} = 3\f{[45]}{\<45\>}\f{[25]\<12\>\<24\>^2[12]^3[24]^2}{\<25\> \<13\>^2\<34\>^2} \\
+ 3\f{[25]\<12\>^2\<24\>[12]^2 [15] [24]^3}{\<25\> \<15\> \<13\>^2\<34\>^2} + \f{[45]}{\<45\>}\f{\<12\>\<14\>\<24\>[12]^3[24]^3}{\<13\>^2\<15\>\<34\>\<35\>} - \f{\left< 12 \right>^2 \<23\>\left< 24 \right>[25] [12]^3 [24]^3 }{\left< 13 \right>^2 \left< 15 \right>\<25\> \left< 34 \right>^2 \left< 35 \right>}\\
- 3\left<45\right>\f{[25]\<12\>^2[12] [15]^2 [24]^3}{\<25\> \<15\> \<13\>^2\<34\>^2} - 9\[45\]\f{[25]\<12\> \<24\> [12]^2 [15] [24]^2}{\<25\> \<13\>^2\<34\>^2}
\end{gathered} 
\ee
Using momentum conservation again in the 4th term in the RHS of the above equation, we finally get
\be
\begin{gathered}
\f{[25]\left< 13 \right>\left< 34 \right> ([13][34])^3}{\lt<25\rt>\left< 12 \right> \left< 15 \right> \left< 24 \right>\left< 54 \right>} + \f{[35]\left< 12 \right> \left< 24 \right> ([12][24])^3 }{\lt<35\rt>\left< 13 \right> \left< 15 \right> \left< 34 \right>\left< 54 \right>} = 3\f{[45]}{\<45\>}\f{[25]\<12\>\<24\>^2[12]^3[24]^2}{\<25\> \<13\>^2\<34\>^2} \\
+ \f{[45]}{\<45\>}\f{\<12\>\<14\>\<24\>[12]^3[24]^3}{\<13\>^2\<15\>\<34\>\<35\>} + 2\f{[25]\<12\>^2\<24\>[12]^2 [15] [24]^3}{\<13\>^2 \<25\> \<15\> \<34\>^2} - \f{[25] \left< 12 \right>^2 \left< 24 \right> [12]^2 [14] [24]^3 }{\left< 13 \right>^2 \left< 15 \right>\<25\> \left< 34 \right> \<35\> }\\
- 3\left<45\right>\f{[25]\<12\>^2[12] [15]^2 [24]^3}{\<25\> \<15\> \<13\>^2\<34\>^2} - 9\[45\]\f{[25]\<12\> \<24\> [12]^2 [15] [24]^2}{\<25\> \<13\>^2\<34\>^2}
\end{gathered} 
\ee
Similarly, the 3rd and 4th terms in RHS of \eqref{sing_terms} give
\be
\begin{gathered}
\f{[24]\left< 13 \right>\left< 35 \right>([13] [35])^3}{\lt<24\rt>\left< 12 \right> \left< 14 \right> \left< 25 \right>\left< 45 \right>} + \f{[34]\left< 12 \right>\left< 25 \right>([12] [25])^3}{\lt<34\rt>\left< 13 \right> \left< 14 \right> \left< 35 \right>\left< 45 \right>} = 3\f{[45]}{\<45\>}\f{[24]\<12\>\<25\>^2[12]^3[25]^2}{\<24\> \<13\>^2\<35\>^2} \\
+ \f{[45]}{\<45\>}\f{\<12\>\<15\>\<25\>[12]^3[25]^3}{\<13\>^2\<14\>\<34\>\<35\>} + 2\f{[24]\<12\>^2\<25\>[12]^2 [14] [25]^3}{\<24\> \<14\> \<13\>^2\<35\>^2} - \f{[24]\left< 12 \right>^2 \left< 25 \right> [12]^2 [15] [25]^3 }{\left< 13 \right>^2 \left< 14 \right>\<24\> \left< 34 \right> \left< 35 \right>}\\
+ 3\left<45\right>\f{[24]\<12\>^2[12] [14]^2 [25]^3}{\<24\> \<14\> \<13\>^2\<35\>^2} + 9\[45\]\f{[24]\<12\> \<25\> [12]^2 [14] [25]^2}{\<24\> \<13\>^2\<35\>^2}
\end{gathered} 
\ee
As is clear from the above equations, we can combine the 12 terms of \eqref{sing_terms} in groups of two as shown above to see that the leading order contribution coming from \eqref{sing_terms} is indeed $\mathcal{O}(\frac{\bar{z}_{45}}{z_{45}})$ instead of the apparent $\mathcal{O}(\frac{1}{z_{45}})$. \\
\\ 
Now, we rewrite the first 4 terms in \eqref{sing_terms} in terms of $ \{\om_i,z_i,\bar z_i\} $, and then expand around $ z_{45} = \bar z_{45} = 0 $. As mentioned earlier, we only keep terms upto $ \mathcal{O}(\bar z_{45}) $ to get
\be
\begin{gathered}
\f{[25]\left< 13 \right>\left< 34 \right> ([13][34])^3}{\lt<25\rt>\left< 12 \right> \left< 15 \right> \left< 24 \right>\left< 54 \right>} + \f{[35]\left< 12 \right> \left< 24 \right> ([12][24])^3 }{\lt<35\rt>\left< 13 \right> \left< 15 \right> \left< 34 \right>\left< 54 \right>} + \f{[24]\left< 13 \right>\left< 35 \right>([13] [35])^3}{\lt<24\rt>\left< 12 \right> \left< 14 \right> \left< 25 \right>\left< 45 \right>} + \f{[34]\left< 12 \right>\left< 25 \right>([12] [25])^3}{\lt<34\rt>\left< 13 \right> \left< 14 \right> \left< 35 \right>\left< 45 \right>}\\
= 2^4 \e_1 \f{\om_1\om_2^4}{\om_3^2\om_4\om_5}\(\om_4 + \om_5\)^3\f{\bar z_{45}}{z_{45}} \f{z_{12}z_{25}\bar z_{12}^3 \bar z_{25}^3}{z_{13}^2 z_{35}^2} \\
- 2^4 \e_1 \f{\om_1\om_2^4}{\om_3^2\om_4 \om_5}\[\(\om_4 + \om_5\)^3 - 5 \om_4\om_5 \( \om_4 + \om_5 \)\]\f{z_{12}^2 \bar z_{12}^2 \bar z_{15} \bar z_{25}^4}{z_{13}^2 z_{15} z_{35}^2}\\
+ 2^4 \e_1 \bar z_{45} \f{z_{12} \bar z_{12}^3 \bar z_{25}^3}{z_{13}^2 z_{15} z_{35}^3} \f{\om_1 \om_2^4}{\om_3^2 \om_4\om_5}\[-z_{15}z_{35} \, \om_4 \(\om_4^2 +6\om_4\om_5 - 3\om_5^2\) + z_{25}z_{35}\(-\om_4^3 + \om_5^3\) \right. \\
\left. + z_{15} z_{25} \(\om_4^3 +6\om_4^2 \om_5 + \om_5^3\)\] + 2^4 \e_1 \bar z_{45} \f{z_{12}^2 \bar z_{12}^2 \bar z_{25}^3}{z_{13}^2 z_{15} z_{35}^2} \f{\om_1 \om_2^4}{\om_3^2 \om_4\om_5}\[\bar z_{25} \, \om_4 \(\om_4^2 - 2\om_5^2\) \right.\\
\left. + \bar z_{15} \(3\om_4^3 - 6\om_4^2 \om_5 - 2\om_4 \om_5^2 + \om_5^3\) \] + \cdots
\end{gathered} 
\ee 
The contribution from the other 8 terms in \eqref{sing_terms} is simply obtained by taking different permutations of 1,2 and 3 in the above expression.
Setting $ \om_4=t\om_P, \ \om_5 = (1-t)\om_P $ and collecting all the singular terms we finally get
\be
B_5^{\tn{S}} = -2^7 \f{\om_P}{t(1-t)} \(\f{\bar z_{45}}{z_{45}} T^{\tn{S}}_L + T^{\tn{S}}_{\mathcal{O}(1)} + \bar z_{45} T^{\tn{S}}_{\bar z}\)
\ee
where
\be
\begin{gathered}
T^{\tn{S}}_L = \[\e_1 \f{z_{12}z_{25}\bar z_{12}^3 \bar z_{25}^3}{z_{13}^2 z_{35}^2}\f{\om_1 \om_2^4 }{\om_3^2} + \e_2 \f{z_{12}z_{15}\bar z_{12}^3 \bar z_{15}^3}{z_{23}^2 z_{35}^2}\f{\om_1^4\om_2 }{\om_3^2} + \e_3 \f{z_{13}z_{15}\bar z_{13}^3 \bar z_{15}^3}{z_{23}^2 z_{25}^2}\f{\om_1^4\om_3 }{\om_2^2}\]
\end{gathered} 
\label{TsL}
\ee
\be
\begin{gathered}
T^{\tn{S}}_{\mathcal{O}(1)} = -\[\e_1 \f{z_{12}^2 \bar z_{12}^2 \bar z_{15} \bar z_{25}^4}{z_{13}^2 z_{15} z_{35}^2} \f{\om_1\om_2^4 }{\om_3^2} +\e_2 \f{z_{12}^2 \bar z_{12}^2 \bar z_{25} \bar z_{15}^4}{z_{23}^2 z_{25} z_{35}^2}\f{\om_1^4\om_2 }{\om_3^2} + \e_3 \f{z_{13}^2 \bar z_{13}^2 \bar z_{35} \bar z_{15}^4}{z_{23}^2 z_{35} z_{25}^2}\f{\om_1^4\om_3 }{\om_2^2}\][1 - 5 t(1-t)]
\end{gathered} 
\label{TsO1}
\ee
and 
\be
\begin{gathered}
T^{\tn{S}}_{\bar z} = \e_1  \f{z_{12} \bar z_{12}^3 \bar z_{25}^3}{z_{13}^2 z_{15} z_{35}^3} \f{\om_1 \om_2^4 }{\om_3^2 }\[-z_{15}z_{35} \, t \(t^2 +6 t(1-t) - 3(1-t)^2\) + z_{25}z_{35}\(-t^3 + (1-t)^3\) \right. \\
\left. + z_{15} z_{25} \(t^3 + 6 t^2 (1-t) + (1-t)^3\)\] + \e_1 \f{z_{12}^2 \bar z_{12}^2 \bar z_{25}^3}{z_{13}^2 z_{15} z_{35}^2} \f{\om_1 \om_2^4 }{\om_3^2}\[\bar z_{25} \, t \(t^2 - 2(1-t)^2\) \right.\\
\left. + \bar z_{15} \(3t^3 - 6t^2 (1-t) - 2t (1-t)^2 + (1-t)^3\) \]\\
+ \e_2 \f{z_{12} \bar z_{12}^3 \bar z_{15}^3}{z_{23}^2 z_{25} z_{35}^3} \f{\om_1^4 \om_2 }{\om_3^2}\[-z_{25}z_{35} \, t \(t^2 +6 t(1-t) - 3(1-t)^2\) + z_{15}z_{35}\(-t^3 + (1-t)^3\) \right. \\
\left. + z_{25} z_{15} \(t^3 +6 t^2 (1-t) + (1-t)^3\)\] + \e_2 \f{z_{12}^2 \bar z_{12}^2 \bar z_{15}^3}{z_{23}^2 z_{25} z_{35}^2} \f{\om_1^4 \om_2 }{\om_3^2 }\[\bar z_{15} \, t \(t^2 - 2(1-t)^2\) \right.\\
\left. + \bar z_{25} \(3 t^3 - 6 t^2 (1-t) - 2t (1-t)^2 + (1-t)^3\) \] \\
+ \e_3 \f{z_{13} \bar z_{13}^3 \bar z_{15}^3}{z_{23}^2 z_{35} z_{25}^3} \f{\om_1^4 \om_3}{\om_2^2}\[-z_{35}z_{25} \, t \(t^2 +6t(1-t) - 3(1-t)^2\) + z_{15}z_{25}\(-t^3 + (1-t)^3\) \right. \\
\left. + z_{35} z_{15} \(t^3 +6 t^2 (1-t) + (1-t)^3\)\] + \e_3 \f{z_{13}^2 \bar z_{13}^2 \bar z_{15}^3}{z_{23}^2 z_{35} z_{25}^2} \f{\om_1^4 \om_3 }{\om_2^2}\[\bar z_{15} \, t \(t^2 - 2(1-t)^2\) \right.\\
\left. + \bar z_{35} \(3t^3 - 6 t^2 (1-t) - 2 t (1-t)^2 + (1-t)^3\) \]
\end{gathered} 
\label{Tsbz}
\ee
\\
Taking into account the other 18 terms (although note that at $\mathcal{O}(\bar{z}_{45})$, only 12 of these contribute and the 9th and 10th term in \eqref{B5_30terms} and the ($1 \leftrightarrow 3$) and ($2 \leftrightarrow 3$) permutation of those do not contribute at this order) in \eqref{B5_30terms} we finally get $ B_5 $ as
\be
B_5 = -2^7 \f{\om_P}{t(1-t)} \(\f{\bar z_{45}}{z_{45}} T_L + T_{\mathcal{O}(1)} + \bar z_{45} T_{\bar z}\) + \cdots
\label{final_b5}
\ee
where we have neglected the higher order terms in the expansion of the RHS of \eqref{B5_30terms} around $ z_{45}=\bar z_{45} = 0 $ in $ \left( \om, z, \bar z \right) $ space and
\be 
\begin{split}
T_L &= T_L^{\tn{S}} + \[\e_1 \e_2 \f{\om_1 \om_2 \om_3^4}{\om_P^3} \f{z_{13} z_{23}\(\bar z_{13} \bar z_{23}\)^3}{z_{15}^2 z_{25}^2} + (1 \leftrightarrow 3) + (2 \leftrightarrow 3)\]\\
T_{\mathcal{O}(1)} &= T^{\tn{S}}_{\mathcal{O}(1)} + \[ \e_1\e_2\e_3 \f{\om_1\om_2\om_P^2}{\om_3} \f{\bar z_{35} (\bar z_{15} \bar z_{25})^3}{z_{35}z_{13}z_{23}} \{(1-t)^5 + t^5\} + \e_1\e_2 \f{\om_P \om_3^4}{\om_1\om_2} \f{\bar z_{12} z_{35}^2 \bar z_{35}^6}{z_{12} z_{15}^2 z_{25}^2} t^2(1-t)^2 \right.\\
&\left. \hspace{6cm} + (1 \leftrightarrow 3) + (2 \leftrightarrow 3)\]\\
T_{\bar z} &= T^{\tn{S}}_{\bar z} + \[ \e_1\e_2 \f{\om_1 \om_2 \om_3^4}{\om_P^3} \f{z_{13} z_{23}\(\bar z_{13} \bar z_{23}\)^3}{z_{15}^2 z_{25}^2}\(\f{1}{z_{15}} + \f{1}{z_{25}}\) + (1 \leftrightarrow 3) + (2 \leftrightarrow 3) \]\\
& - \[ \e_1\e_2\e_3 \f{\om_1\om_2 \om_P^2}{\om_3} \f{ (\bar z_{15} \bar z_{25})^3}{z_{13}z_{23} z_{35}} (1-t)^5 + 3\e_1\e_2\e_3 \f{\om_1\om_2 \om_P^2}{\om_3} \f{\bar z_{35}(\bar z_{15} \bar z_{25} )^2 (\bar z_{15} + \bar z_{25} )}{ z_{13} z_{23} z_{35}} t^5 \right.\\
& \left. \hspace{4cm} + \ \e_1\e_2 \f{\om_3^4 \om_P}{\om_1\om_2} \f{\bar z_{12} z_{35}^2 \bar z_{35}^5}{z_{12} z_{15}^2 z_{25}^2} t^2(1-t)^2 + (1 \leftrightarrow 3) + (2 \leftrightarrow 3) \]
\end{split}
\label{expl_Ts}
\ee
\\
We will now Mellin transform \eqref{final_b5} and take the OPE limit $ 4 \to 5 $. We want to emphasize here that after Mellin transformation, the higher order terms in the OPE expansion of the Mellin amplitude may receive contribution from the lower order terms. This is because of the fact that, after Mellin transformation the Mellin amplitude will depend on $ \tilde{\om}^*_i $'s  as well as the delta function involving cross ratios coming from 5-point momentum conserving delta function as parametrized in \eqref{5pt_delta_fn}. In the next subsection we analyse this in detail and calculate the order by order terms in the OPE expansion $ 4\to 5 $ of the 5-point Mellin amplitude.
\section{Mellin Transformation of the 5-Point Amplitude}\label{mellin_5_pt}

For the discussion of this Appendix, the prefactor $ \f{i}{(4\pi)^2 960} $ in \eqref{5pt_sdg_b5} is not important. Thus we only Mellin transform $ B_5 $ and keep terms only upto $ \mathcal{O}(\bar z_{45}) $. Substituting \eqref{final_b5} in \eqref{mellin_trans_b5} we get:
\bea
\no \widetilde{B}_5 = -2^7\int_0^\infty \prod_{i=1}^5 d\om_i \, \om_i^{\D_i-1} e^{-i\sum_{i=1}^4 \e_i \om_i u_{i5}} \f{\om_P}{t(1-t)} \(\f{\bar z_{45}}{z_{45}} T_L(\om_1,\om_2,\om_3,\om_P) + T_{\mathcal{O}(1)}(\om_1,\om_2,\om_3,\om_P) \right.\\
\left. + \bar z_{45} T_{\bar z}(\om_1,\om_2,\om_3,\om_P)\) \d^{(4)}\(\sum_{i=1}^5 \e_i \om_i q_i \) \ \ \ \ \label{mellin_trans_b5_app}
\eea
where $ T_L, \ T_{\mathcal{O}(1)} $ and $ \ T_{\bar z} $ are given by \eqref{expl_Ts} and we have kept their $ \{\om \} $ dependence explicit for our convenience. Also we have used momentum conservation in the exponential. Now using the parametrization \eqref{5pt_delta_fn}, we can perform the $ (\om_1,\om_2,\om_3) $ integrals to obtain:
\be
\begin{gathered}
 \widetilde{B}_5 = -2^5 \int_0^1 dt \, t^{\D_4-2}(1-t)^{\D_5-2} \int_0^\infty d\om_P \, \om_P^{\D_4+\D_5-1}\prod_{i=1}^3 (\om^*_i)^{\D_i-1} e^{-i\sum_{i=1}^3 \e_i \om^*_i u_{i5} -i\om_P t \, u_{45}}  \\
 \times \(\f{\bar z_{45}}{z_{45}} T_L(\om^*_1,\om^*_2,\om^*_3,\om_P) + T_{\mathcal{O}(1)}(\om^*_1,\om^*_2,\om^*_3,\om_P) + \bar z_{45} T_{\bar z}(\om^*_1,\om^*_2,\om^*_3,\om_P)\)  \\
 \times \d\( x - \bar x - t z_{45}\(\f{x}{z_{35}}-\f{\bar x}{z_{25}}\) - t \bar z_{45}\(\f{x}{\bar z_{25}}-\f{\bar x}{\bar z_{35}}\) + t z_{45} \bar z_{45}\(\f{x}{z_{35}\bar z_{25}}-\f{\bar x}{z_{25}\bar z_{35}}\) \)
\end{gathered} 
\label{mellin_trans_b5_app_1}
\ee
Now from  \eqref{om_st} and the explicit expressions of $ T_L, \ T_{\mathcal{O}(1)} $ and $ T_{\bar z} $ given by \eqref{expl_Ts} one can see that
\be 
\begin{split}
T_L(\om^*_1,\om^*_2,\om^*_3,\om_P) &= \om_P^3 \, \mathcal{T}_L(\tilde{\om}^*_1, \tilde{\om}^*_2, \tilde{\om}^*_3)\\
T_{\mathcal{O}(1)}(\om^*_1,\om^*_2,\om^*_3,\om_P) &= \om_P^3 \, \mathcal{T}_{\mathcal{O}(1)}(\tilde{\om}^*_1, \tilde{\om}^*_2, \tilde{\om}^*_3)\\
 T_{\bar z}(\om^*_1,\om^*_2,\om^*_3,\om_P) &= \om_P^3 \, \mathcal{T}_{\bar z}(\tilde{\om}^*_1, \tilde{\om}^*_2, \tilde{\om}^*_3)
\end{split}
\ee
where 
\be 
\begin{split}
\mathcal{T}_L(\tilde{\om}^*_1, \tilde{\om}^*_2, \tilde{\om}^*_3) &= \mathcal{T}_L^{\tn{S}}(\tilde{\om}^*_1, \tilde{\om}^*_2, \tilde{\om}^*_3) + \[\e_1 \e_2 \, \tilde{\om}^*_1 \tilde{\om}^*_2 (\tilde{\om}^*_3)^4 \f{z_{13} z_{23}\(\bar z_{13} \bar z_{23}\)^3}{z_{15}^2 z_{25}^2} + (1 \leftrightarrow 3) + (2 \leftrightarrow 3)\]\\
\mathcal{T}_{\mathcal{O}(1)}(\tilde{\om}^*_1, \tilde{\om}^*_2, \tilde{\om}^*_3) &= \mathcal{T}^{\tn{S}}_{\mathcal{O}(1)}(\tilde{\om}^*_1, \tilde{\om}^*_2, \tilde{\om}^*_3) + \[ \e_1\e_2\e_3 \f{\tilde{\om}^*_1 \tilde{\om}^*_2 }{\tilde{\om}^*_3} \f{\bar z_{35} (\bar z_{15} \bar z_{25})^3}{z_{35}z_{13}z_{23}} \{(1-t)^5 + t^5\} \right.\\
&\left. \hspace{3cm} + \e_1\e_2 \f{(\tilde{\om}^*_3)^4}{\tilde{\om}^*_1 \tilde{\om}^*_2} \f{\bar z_{12} z_{35}^2 \bar z_{35}^6}{z_{12} z_{15}^2 z_{25}^2} t^2(1-t)^2 + (1 \leftrightarrow 3) + (2 \leftrightarrow 3)\]\\
\mathcal{T}_{\bar z}(\tilde{\om}^*_1, \tilde{\om}^*_2, \tilde{\om}^*_3) &= \mathcal{T}^{\tn{S}}_{\bar z}(\tilde{\om}^*_1, \tilde{\om}^*_2, \tilde{\om}^*_3) + \[ \e_1\e_2 \, \tilde{\om}^*_1 \tilde{\om}^*_2 (\tilde{\om}^*_3)^4 \f{z_{13} z_{23}\(\bar z_{13} \bar z_{23}\)^3}{z_{15}^2 z_{25}^2}\(\f{1}{z_{15}} + \f{1}{z_{25}}\) + (1 \leftrightarrow 3) + (2 \leftrightarrow 3) \]\\
& - \[ \e_1\e_2\e_3 \f{\tilde{\om}^*_1 \tilde{\om}^*_2 }{\tilde{\om}^*_3} \f{ (\bar z_{15} \bar z_{25})^3}{z_{13}z_{23} z_{35}} (1-t)^5 + 3\e_1\e_2\e_3 \f{\tilde{\om}^*_1 \tilde{\om}^*_2 }{\tilde{\om}^*_3} \f{\bar z_{35}(\bar z_{15} \bar z_{25} )^2 (\bar z_{15} + \bar z_{25} )}{ z_{13} z_{23} z_{35}} t^5 \right.\\
& \left. \hspace{4cm} + \ \e_1\e_2 \f{(\tilde{\om}^*_3)^4}{\tilde{\om}^*_1 \tilde{\om}^*_2} \f{\bar z_{12} z_{35}^2 \bar z_{35}^5}{z_{12} z_{15}^2 z_{25}^2} t^2(1-t)^2 + (1 \leftrightarrow 3) + (2 \leftrightarrow 3) \]
\end{split}
\label{expl_curlyTs}
\ee
and $ \mathcal{T}_L^{\tn{S}}(\tilde{\om}^*_1, \tilde{\om}^*_2, \tilde{\om}^*_3), \mathcal{T}^{\tn{S}}_{\mathcal{O}(1)}(\tilde{\om}^*_1, \tilde{\om}^*_2, \tilde{\om}^*_3) $ and $ \mathcal{T}^{\tn{S}}_{\bar z}(\tilde{\om}^*_1, \tilde{\om}^*_2, \tilde{\om}^*_3) $ are given by \eqref{TsL}, \eqref{TsO1} and \eqref{Tsbz} respectively with $ \{\om_1,\om_2,\om_3\} $ replaced by $ \{ \tilde{\om}^*_1, \tilde{\om}^*_2, \tilde{\om}^*_3 \} $. Now we can perform the $ \om_P $ integral in \eqref{mellin_trans_b5_app_1} and obtain:
\be
\begin{gathered}
 \widetilde{B}_5 = -2^5 \G(\D) \int_0^1 dt \, t^{\D_4-2}(1-t)^{\D_5-2} \f{ \prod_{i=1}^3 (\tilde{\om}^*_i)^{\D_i-1}}{\left[ i \( \sum_{i=1}^3 \e_i \tilde{\om}^*_i u_{i5} + t \, u_{45}\) \right]^\D} \\
 \times \(\f{\bar z_{45}}{z_{45}} \mathcal{T}_L(\tilde{\om}^*_1, \tilde{\om}^*_2, \tilde{\om}^*_3) + \mathcal{T}_{\mathcal{O}(1)}(\tilde{\om}^*_1, \tilde{\om}^*_2, \tilde{\om}^*_3) + \bar z_{45} \mathcal{T}_{\bar z}(\tilde{\om}^*_1, \tilde{\om}^*_2, \tilde{\om}^*_3)\)  \\
 \times \d\( x - \bar x - t z_{45}\(\f{x}{z_{35}}-\f{\bar x}{z_{25}}\) - t \bar z_{45}\(\f{x}{\bar z_{25}}-\f{\bar x}{\bar z_{35}}\) + t z_{45} \bar z_{45}\(\f{x}{z_{35}\bar z_{25}}-\f{\bar x}{z_{25}\bar z_{35}}\) \)
\end{gathered} 
\label{mellin_trans_b5_app_2}
\ee
where $ \D = \sum_{i=1}^5 \D_i $. We now expand the above equation around $ z_{45} = \bar z_{45} = u_{45} = 0 $.

\subsection{Evaluating the Leading Order Contribution}

It is clear from \eqref{mellin_trans_b5_app_2} that the leading order term goes as $ \sim \f{\bar z_{45}}{z_{45}} $ and the contribution to the leading order can come only from the term containing $ \mathcal{T}_L(\tilde{\om}^*_1, \tilde{\om}^*_2, \tilde{\om}^*_3) $. At leading order we have $ \tilde{\om}^*_i = \e_i \s_{i,1} $. Thus, the leading order term of $ \widetilde{B}_5 $ is given by:
\be
\begin{gathered}
 \widetilde{B}_5|_{\mathcal{O}(\frac{\bar{z}_{45}}{z_{45}})} = -2^5 \f{\bar z_{45}}{z_{45}}\f{ \G(\D)}{\left( i \mathcal{D} \right)^\D} \prod_{i=1}^3 (\e_i \s_{i,1})^{\D_i-1} \int_0^1 dt \, t^{\D_4-2}(1-t)^{\D_5-2}  \mathcal{T}_L(\e_1\s_{1,1}, \e_2\s_{2,1}, \e_3\s_{3,1}) \d\( x - \bar x  \)
\end{gathered} 
\label{lead_ord_app_1}
\ee
where $ \mathcal{D} = \( \sum_{i=1}^3 \s_{i,1} u_{i5} \) $. From \eqref{TsL} and the first equation of \eqref{expl_Ts} we have
\be 
\begin{gathered}
\mathcal{T}_L(\e_1\s_{1,1}, \e_2\s_{2,1}, \e_3\s_{3,1}) =  \[ \f{z_{12}z_{25}\bar z_{12}^3 \bar z_{25}^3}{z_{13}^2 z_{35}^2}\f{\s_{1,1} \s_{2,1}^4 }{\s_{3,1}^2} + \f{z_{12}z_{15}\bar z_{12}^3 \bar z_{15}^3}{z_{23}^2 z_{35}^2}\f{\s_{1,1}^4\s_{2,1} }{\s_{3,1}^2} + \f{z_{13}z_{15}\bar z_{13}^3 \bar z_{15}^3}{z_{23}^2 z_{25}^2}\f{\s_{1,1}^4\s_{3,1} }{\s_{2,1}^2}\]\\
+ \[ \s_{1,1} \s_{2,1} \s_{3,1}^4 \f{z_{13} z_{23}\(\bar z_{13} \bar z_{23}\)^3}{z_{15}^2 z_{25}^2} + (1 \leftrightarrow 3) + (2 \leftrightarrow 3)\]
\end{gathered}
\label{TauL}
\ee
Now, using \eqref{sig11}-\eqref{xxb}, one can show that
\be
\begin{split}
 \f{z_{12}z_{25}\bar z_{12}^3 \bar z_{25}^3}{z_{13}^2 z_{35}^2}\f{\s_{1,1} \s_{2,1}^4 }{\s_{3,1}^2} &= \f{z_{12}z_{13} (\bar z_{12} \bar{z}_{13})^3}{z_{25}^2 z_{35}^2}\s_{1,1}^4 \s_{2,1}\s_{3,1} \\
 \f{z_{12}z_{15}\bar z_{12}^3 \bar z_{15}^3}{z_{23}^2 z_{35}^2}\f{\s_{1,1}^4\s_{2,1} }{\s_{3,1}^2} &= \f{z_{12}z_{23} (\bar z_{12} \bar{z}_{23})^3}{z_{15}^2 z_{35}^2}\s_{1,1} \s_{2,1}^4\s_{3,1} \\
 \f{z_{13}z_{15}\bar z_{13}^3 \bar z_{15}^3}{z_{23}^2 z_{25}^2}\f{\s_{1,1}^4\s_{3,1} }{\s_{2,1}^2} &= \f{z_{13}z_{23} (\bar z_{13} \bar{z}_{23})^3}{z_{15}^2 z_{25}^2}\s_{1,1} \s_{2,1} \s_{3,1}^4
 \end{split}
\ee
Using the above relations, we can simplify \eqref{TauL} to get
\begin{align}
\no \mathcal{T}_L(\e_1\s_{1,1}, \e_2\s_{2,1}, \e_3\s_{3,1}) &= 2 \[ \s_{1,1} \s_{2,1} \s_{3,1}^4 \f{z_{13} z_{23}\(\bar z_{13} \bar z_{23}\)^3}{z_{15}^2 z_{25}^2} + (1 \leftrightarrow 3) + (2 \leftrightarrow 3)\] \\ &= 2 \[ \mathcal{N}_4 + \mathcal{N}_4(1 \leftrightarrow 3) + \mathcal{N}_4(2 \leftrightarrow 3)\]
\end{align}
where the second equality follows from \eqref{defN_4}. Since this is independent of $t$, we can easily carry out the $t$-integral in \eqref{lead_ord_app_1} to get
\begin{align}
\no \tilde{B}_5|_{\mathcal{O}(\frac{\bar{z}_{45}}{z_{45}})} &= -2^6 \f{\bar z_{45}}{z_{45}}B(\Delta_4 - 1,\Delta_5 - 1)\f{\G(\Delta)}{\(i\mathcal{D}\)^{\Delta}}\d(x-\bar x) \\ &\times \prod_{i=1}^3 (\e_i\s_{i,1})^{\Delta_i - 1} \[ \mathcal{N}_4 + \mathcal{N}_4\(1\leftrightarrow 3\) + \mathcal{N}_4\(2\leftrightarrow 3\) \] \label{lorder_final}
\end{align}
This precisely gives the equation \eqref{lead_B5}.



\subsection{Evaluating the $ \mathcal{O}(1) $ Contribution}
From \eqref{mellin_trans_b5_app_2}, we can see that the $  \mathcal{O}(1) $ contribution to the 5-point amplitude essentially comes only from the term containing $ \mathcal{T}_{\mathcal{O}(1)}(\tilde{\om}^*_1,\tilde{\om}^*_2,\tilde{\om}^*_3) $ when $ \tilde{\om}^*_i $'s take their leading order value given by $ \e_i \s_{i,1} $. Let us write the Mellin integral at order 1:
\be
\begin{gathered}
 \widetilde{B}_5\big|_{\mathcal{O}(1)} = -2^5 \f{ \G(\D)}{\left( i \mathcal{D} \right)^\D} \prod_{i=1}^3 (\e_i \s_{i,1})^{\D_i-1} \int_0^1 dt \, t^{\D_4-2}(1-t)^{\D_5-2}  \mathcal{T}_{\mathcal{O}(1)}(\e_1\s_{1,1}, \e_2\s_{2,1}, \e_3\s_{3,1}) \d\( x - \bar x  \)
\end{gathered} 
\label{ord_1_app_1}
\ee
 We will not attempt to take the explicit expressions of $ \mathcal{T}_{\mathcal{O}(1)}(\e_1 \s_{1,1},\e_2 \s_{2,1},\e_3 \s_{3,1}) $ and Mellin integrate it. Rather we will take a different approach which is more helpful for our purpose of the OPE factorization. Firstly, from the second equation of \eqref{expl_Ts} we observe that $ \mathcal{T}_{\mathcal{O}(1)}(\e_1 \s_{1,1},\e_2 \s_{2,1},\e_3 \s_{3,1}) $ is a polynomial in $ t $ with the highest power being $ 4 $. We use this fact and write $ \mathcal{T}_{\mathcal{O}(1)}(\e_1 \s_{1,1},\e_2 \s_{2,1},\e_3 \s_{3,1}) $ as
\be
 \mathcal{T}_{\mathcal{O}(1)}(\e_1 \s_{1,1},\e_2 \s_{2,1},\e_3 \s_{3,1}) = \sum_{k=0}^4 t^k \mathcal{F}^{(1)}_{k}(\{\e_i, z_i,\bar z_i\}) 
 \label{order1_t4}
\ee
The explicit expressions for the functions $ \mathcal{F}^{(1)}_{k}(\{\e_i, z_i,\bar z_i\}) $ can be read out from the second equation of \eqref{expl_Ts}. However, they are not relevant for our discussions and hence we will not write them explicitly. Using \eqref{order1_t4}, we can easily evaluate the integral \eqref{ord_1_app_1} to get,
\be
\begin{gathered}
 \widetilde{B}_5\big|_{\mathcal{O}(1)} = -2^5 \f{ \G(\D)}{\left( i \mathcal{D} \right)^\D} \prod_{i=1}^3 (\e_i \s_{i,1})^{\D_i-1} \sum_{k=0}^4 B(\D_4+k-1,\D_5-1) \mathcal{F}^{(1)}_{k}(\{\e_i, z_i,\bar z_i\}) \d\( x - \bar x  \)
\end{gathered} 
\label{ord_1_app_2}
\ee
This is the expression we have used in section \eqref{sec:O1}.
\subsection{Evaluating the Order $ \mathcal{O}(\bar z_{45}) $ Contribution}
We apply the same strategy as the previous section here. However, we have to be careful now as there will be contributions at  $ \mathcal{O}(\bar z_{45}) $ from the lower order terms. Like $ \mathcal{O}(1) $ terms, here also we are only concerned about the $ t $-dependence. Before proceeding further let us first write down the expansion of different components in \eqref{mellin_trans_b5_app_2} around $ z_{45} = \bar z_{45} = u_{45} = 0 $. Keeping terms only upto $ \mathcal{O}(\bar z_{45}) $ we have
\be
\begin{split}
\tilde{\om}_i^* &= \e_i \(\s_{i,1} + t z_{45} \s_{i,2} + t \bar z_{45}\s_{i,3} \)\\
 \mathcal{T}_L(\tilde{\om}^*_1, \tilde{\om}^*_2, \tilde{\om}^*_3) &= \mathcal{T}_L(\e_1 \s_{1,1}, \e_2 \s_{2,1}, \e_3\s_{3,1}) + z_{45} \mathcal{T}_L^{(z)}(\{\e_i,z_i,\bar z_i\}) + \bar z_{45} \mathcal{T}_L^{(\bar z)}(\{\e_i,z_i,\bar z_i\}) \\
\mathcal{T}_{\mathcal{O}(1)}(\tilde{\om}^*_1, \tilde{\om}^*_2, \tilde{\om}^*_3) &= \mathcal{T}_{\mathcal{O}(1)}(\e_1 \s_{1,1}, \e\s_{2,1}, \e_3\s_{3,1}) + z_{45} \mathcal{T}_{\mathcal{O}(1)}^{(z)}(\{\e_i,z_i,\bar z_i\}) + \bar z_{45} \mathcal{T}_{\mathcal{O}(1)}^{(\bar z)}(\{\e_i,z_i,\bar z_i\}) \\
\mathcal{T}_{\bar z}(\tilde{\om}^*_1, \tilde{\om}^*_2, \tilde{\om}^*_3) &= \mathcal{T}_{\bar z}(\e_1 \s_{1,1}, \e_2 \s_{2,1}, \e_3 \s_{3,1}) + z_{45} \mathcal{T}_{\bar z}^{(z)}(\{\e_i,z_i,\bar z_i\}) + \bar z_{45} \mathcal{T}_{\bar z}^{(\bar z)}(\{\e_i,z_i,\bar z_i\})
\end{split} 
\ee
The explicit expressions for different $ \mathcal{T} $'s are not required for our discussions. For notational convenience, we will not write the arguments of different $ \mathcal{T} $'s and replace $ \mathcal{T}_{L,\mathcal{O}(1),\bar z}(\e_1 \s_{1,1}, \e_2 \s_{2,1}, \e_3 \s_{3,1}) $ by $ \mathcal{T}^{(0)}_{L,\mathcal{O}(1),\bar z} $. Let us first write down all possible contributions to $ \widetilde{B}_5 $ at $ \mathcal{O}(\bar z_{45}) $. From \eqref{mellin_trans_b5_app_2} we have,
\be
\begin{gathered}
 \widetilde{B}_5\big|_{\mathcal{O}(\bar z_{45})} = -2^5 \f{\G(\D)}{\(i\mathcal{D}\)^\D} \prod_{i=1}^3 (\e_i \s_{i,1})^{\D_i-1} \int_0^1 dt \, t^{\D_4-2}(1-t)^{\D_5-2} \[ \( \mathcal{T}^{(z)}_L + \mathcal{T}^{(\bar z)}_{\mathcal{O}(1)} + \mathcal{T}^{(0)}_{\bar z}\)\d\( x - \bar x\) \right.\\
\left. - t \left\{ \(\f{x}{z_{35}}-\f{\bar x}{z_{25}}\)\mathcal{T}^{(0)}_L + \mathcal{T}^{(0)}_{\mathcal{O}(1)} \(\f{x}{\bar z_{25}} -\f{\bar x}{\bar z_{35}}\) \right\} \d'\( x - \bar x\)\right.\\
\left. + t \left\{ (\D_1-1) \f{\s_{1,2}}{\s_{1,1}} + (\D_2-1) \f{\s_{2,2}}{\s_{2,1}} + (\D_3-1) \f{\s_{3,2}}{\s_{3,1}} - \D \f{\sum_{i=1}^3 \s_{i,2}u_{i5}}{\mathcal{D}} \right\} \mathcal{T}^{(0)}_L\d\( x - \bar x\) \right.\\
\left. + t \left\{ (\D_1-1) \f{\s_{1,3}}{\s_{1,1}} + (\D_2-1) \f{\s_{2,3}}{\s_{2,1}} + (\D_3-1) \f{\s_{3,3}}{\s_{3,1}} - \D \f{\sum_{i=1}^3 \s_{i,3}u_{i5}}{\mathcal{D}} \right\}
  \mathcal{T}^{(0)}_{\mathcal{O}(1)}\d\( x - \bar x\) \]
\end{gathered} 
\label{order_bz_app_1}
\ee

Now, by expanding the $ \mathcal{T} $'s in \eqref{expl_Ts} around $ z_{45} = \bar z_{45} = 0 $ and keeping terms only upto $ \mathcal{O}(\bar z_{45}) $, one can check that all the terms at different orders in the expansion are polynomial of $ t $. The highest degree of polynomial is $ 5 $ and appears in $ \mathcal{T}^{(0)}_{\bar z} $ only. All the other $ \mathcal{T} $'s have less power of $ t $. Thus we conclude that the terms in the parenthesis $ [\cdots] $ in \eqref{order_bz_app_1} can be written as a polynomial of $ t $ in the following way
\be 
\begin{gathered}
\[ \( \mathcal{T}^{(z)}_L + \mathcal{T}^{\bar z}_{\mathcal{O}(1)} + \mathcal{T}^{(0)}_{\bar z}\)\d\( x - \bar x\) - t \left\{ \(\f{x}{z_{35}}-\f{\bar x}{z_{25}}\)\mathcal{T}^{(0)}_L + \mathcal{T}^{(0)}_{\mathcal{O}(1)} \(\f{x}{\bar z_{25}} -\f{\bar x}{\bar z_{35}}\) \right\} \d'\( x - \bar x\)\right.\\
\left. + t \left\{ (\D_1-1) \f{\s_{1,2}}{\s_{1,1}} + (\D_2-1) \f{\s_{2,2}}{\s_{2,1}} + (\D_3-1) \f{\s_{3,2}}{\s_{3,1}} - \D \f{\sum_{i=1}^3 \s_{i,2}u_{i5}}{\mathcal{D}} \right\} \mathcal{T}^{(0)}_L\d\( x - \bar x\) \right.\\
\left. + t \left\{ (\D_1-1) \f{\s_{1,3}}{\s_{1,1}} + (\D_2-1) \f{\s_{2,3}}{\s_{2,1}} + (\D_3-1) \f{\s_{3,3}}{\s_{3,1}} - \D \f{\sum_{i=1}^3 \s_{i,3}u_{i5}}{\mathcal{D}} \right\}
  \mathcal{T}^{(0)}_{\mathcal{O}(1)}\d\( x - \bar x\) \]\\
  = \sum_{k=1}^5 t^k \mathcal{F}^{(\bar z)}_k (\{\e_i, z_i,\bar z_i\})
\end{gathered}
\label{eq:771}
\ee
where once again the explicit expressions of $ \mathcal{F}^{(\bar z)}_k (\{\e_i, z_i,\bar z_i\}) $ are not relevant for our discussions. Substituting \eqref{eq:771} in \eqref{order_bz_app_1} and performing the $ t $-integral, we finally get:
\be
\begin{gathered}
 \widetilde{B}_5\big|_{\mathcal{O}(\bar z_{45})} = -2^5 \f{\G(\D)}{\(i\mathcal{D}\)^\D} \prod_{i=1}^3 (\e_i \s_{i,1})^{\D_i-1} \sum_{k=1}^5 B(\D_4+k-1,\D_5-1) \mathcal{F}^{(\bar z)}_k (\{\e_i, z_i,\bar z_i\})
\end{gathered} 
\label{order_bz_app_2}
\ee
This is the form for the $\mathcal{O}(\bar{z}_{45})$ 5-point amplitude which we use in the main text of this paper.
\section{$ w $-Algebra Primaries}
\label{wap}




Let's start with the universal term in the OPE between two positive helicity hard gravitons given by,
\be
G^+_{\D_1}(z,\bar z)G^+_\D(0,0) = -\f{1}{z} \sum_{n=0}^\infty B(\D_1-1+n,\D-1) \f{\bar z^{n+1}}{n!}\bar \pa^n G^+_{\D+\D_1}(0,0)
\ee
We now take the conformal soft limit, first by setting $ \D_1=k+\varepsilon $ and then taking $ \varepsilon \to 0 $ to get,
\be
\begin{gathered}
\lim_{\varepsilon \to 0} \varepsilon G^+_{k+\varepsilon}(z,\bar z)G^+_\D(0,0) = -\f{1}{z} \sum_{n=0}^\infty \[\lim_{\varepsilon \to 0} \varepsilon B(k-1+n+\varepsilon,\D-1)\] \f{\bar z^{n+1}}{n!} \bar \pa^n G^+_{\D+k}(0,0)\\
\Rightarrow H^k(z,\bar z)G^+_\D(0,0) = -\f{1}{z} \sum_{n=0}^\infty \[\lim_{\varepsilon \to 0} \varepsilon B(k-1+n+\varepsilon,\D-1)\] \f{\bar z^{n+1}}{n!} \bar \pa^n G^+_{\D+k}(0,0)\\
\end{gathered}
\ee
Next, we mode expand the soft graviton operator $ H^k(z,\bar z) $ on the LHS of the above equation according to \eqref{mode1} and get, 
\be\label{F3}
\begin{gathered}
\sum_{m=\f{k-2}{2}}^{\f{2-k}{2}} \f{H^k_m(z)}{\bar z^{m+\f{k-2}{2}} } G^+_\D(0,0) = -\f{1}{z} \sum_{n=0}^\infty \[\lim_{\varepsilon \to 0} \varepsilon B(k-1+n+\varepsilon,\D-1)\] \f{\bar z^{n+1}}{n!} \bar \pa^n G^+_{\D+k}(0,0)
\end{gathered}
\ee
By comparing the terms at order $ \bar z^{n+1} $ on both the sides of the above equation for $0\leq n \leq 1-k $, we get,  
\be
\begin{gathered}
 H^k_{\f{2-k}{2}-n-1}(z) G^+_\D(0,0) = -\f{1}{z} \[\lim_{\varepsilon \to 0} \varepsilon B(k-1+n+\varepsilon,\D-1)\] \f{1}{n!}\bar \pa^n G^+_{\D+k}(0,0)
\end{gathered}
\ee
Now we use the holomorphic mode expansion \eqref{mode2} of the currents $   H^k_{\f{2-k}{2}-n-1}(z) $ in the above equation and obtain,
\be\label{eq:F5}
\begin{gathered}
\sum_{\a} z^{-\a-\f{k+2}{2}}H^k_{\a,\f{2-k}{2}-n-1} G^+_\D(0,0) = -\f{1}{z} \[\lim_{\varepsilon \to 0} \varepsilon B(k-1+n+\varepsilon,\D-1)\] \f{1}{n!}\bar \pa^n G^+_{\D+k}(0,0)
\end{gathered}
\ee
We can see from the above equation, that there is only a simple pole at $ z=0 $ on the RHS. Thus, the holomorphic singularity structure of the above equation \eqref{eq:F5} tells us that the following conditions should hold,
\be
\begin{gathered}
 H^k_{-\f{k+2}{2}+m,\f{2-k}{2}-n-1} G^+_\D(0,0) = - \[\lim_{\varepsilon \to 0} \varepsilon B(k-1+n+\varepsilon,\D-1)\] \f{1}{n!} \bar \pa^n G^+_{\D+k}(0,0)
 \end{gathered}
\ee
for $m = 1$ and 
\be
\begin{gathered}
 H^k_{-\f{k+2}{2}+m,\f{2-k}{2}-n-1} G^+_\D(0,0) = 0
\end{gathered}
\ee
for $m > 1$ and $0\leq n \leq 1-k$ with $k = 1,0,-1, \cdots$. \\
Moreover, from \eqref{F3}, one can see that there is no term on the RHS that goes like $ \bar z^0 $. Thus on the LHS, the coefficients of the $ \bar z^0 $ term should also vanish which gives the following condition,
\be
\begin{gathered}
H^k_{\f{2-k}{2}}(z) G^+_\D(0,0) = 0
\label{modeq}
\end{gathered} 
\ee
This equation implies
\be
H^k_{-\f{k+2}{2}+m,\f{2-k}{2}} G^+_\D(0,0) = 0, \ m\geq 1 
\ee

\section{Transformation of the MHV Null States under $ {sl_2(R)}_V $ and $ \overline{sl_2(R)} $ Algebras}
\label{mnt}

In this section of the Appendix, we list the transformation properties of all the MHV null states appearing at different orders of the OPE between two positive helicity outgoing gravitons under $ {sl_2(R)}_V $ and $ \overline{sl_2(R)} $ algebras. Let us first write down their explicit expressions in terms of the descendants of the $ w $-algebra. We first write down the actions of the $ H^{-1}_{\f{1}{2},\f{1}{2}} $ on the null states $ \Phi_k(\D) $ given by \eqref{phn} and $ \Psi_k(\D) $ given by  \eqref{psn}. They are given by
\be \label{nt1}
\begin{split}
H^{-1}_{\f{1}{2},\f{1}{2}}\Phi_k(\D) &= -\f{1}{2}(k+1)(k+2)\Phi_{k+1}(\D-1) -\f{1}{2}(\D+k-3)(\D+k-4)\Phi_{k}(\D-1)\\
& + \f{(-1)^k}{k!}\f{\G(\D+k-2)}{\G(\D-2)}\Phi_1(\D-1) \\
H^{-1}_{\f{1}{2},\f{1}{2}}\Psi_k(\D) &= -\f{1}{2}(k+2)(k-1)\Psi_{k+1}(\D-1) - \f{1}{2}(\D+k-3)(\D+k-4)\Psi_k(\D-1)\\
& - \f{(-1)^k}{k!}\f{\G(\D+k-2)}{\G(\D-2)} \Psi_1(\D-1)\\
H^{-1}_{\f{1}{2},\f{1}{2}}\Omega_k(\D) &=\f{1}{2}(k+1)(k+2)\Omega_k(\D-1) - \f{1}{2}(\D-4)(\D-5)\Omega_k(\D-1) \\ 
& - \f{1}{2}(k+1)(k+2)\Omega_{k+1}(\D-1)\\
H^{-1}_{\f{1}{2},\f{1}{2}} \Pi_k(\D) &= \f{1}{2}k(k+1)\Pi_k(\D-1) -\f{1}{2}(\D-4)(\D-5)\Pi_k(\D-1) \\ 
& -\f{1}{2}(k-1)(k+2)\Pi_{k+1}(\D-1)
\end{split}
\ee
The actions of $ H^{1}_{-\f{1}{2},-\f{1}{2}} $ on the MHV null states are given by,
\be\label{nt2}
\begin{split}
H^{1}_{-\f{1}{2},-\f{1}{2}} \Phi_k(\D) &= -\Phi_{k}(\D+1)-\Phi_{k-1}(\D+1)\\
H^{1}_{-\f{1}{2},-\f{1}{2}} \Psi_k(\D) &= -\Psi_{k}(\D+1)-\Psi_{k-1}(\D+1)\\
H^{1}_{-\f{1}{2},-\f{1}{2}}\Omega_k(\D) &= - \Omega_k(\D+1)\\
H^{1}_{-\f{1}{2},-\f{1}{2}} \Pi_k(\D) &= - \Pi_{k+1}(\D+1) 
\end{split}
\ee
The actions of $ H^{0}_{0,1} $ on the MHV null states are given by,
\be\label{nt3}
\begin{split}
H^0_{0,1} \Phi_k(\D) &= 0 \\
H^0_{0,1} \Psi_k(\D) &=  (k+2)\Phi_{k+1}(\D-1)-2\f{(-1)^k}{k!}\f{\G(\D+k-2)}{\G(\D-2)}\Phi_1(\D-1)\\
H^0_{0,1} \Omega_k(\D) &= 0 \\
H^{0}_{0,1} \Pi_{k}(\D) &= - (\D+k-3)\Omega_{k}(\D-1) + (k+2) \Omega_{k+1}(\D-1)
\end{split} 
\ee
In deriving the above transformation properties, we have used the algebra \eqref{hsa} and the action of different operators on the primaries given in Appendix \ref{wap}.

\section{Review of General Structure of $ w $-Invariant OPE}
\label{revgs}

It was shown in \cite{Banerjee:2023zip}, that the OPE between two positive helicity outgoing graviton primaries of any $ w $-invariant  theory can always be written in terms of the MHV OPE's and its null states. The MHV null states that can appear at $\mathcal{O}({z^0\bar z^0})$ and $\mathcal{O}({z^0\bar z})$ are given by \cite{Banerjee:2020zlg,Banerjee:2021cly}
\be\label{phn}
\Phi_k(\D) = \[ H^{1-k}_{\f{k-3}{2},\f{k+1}{2}} \(-H^{1}_{-\f{1}{2},-\f{1}{2}}\)^k -\f{(-1)^k}{k!} \f{\G(\D +k-2)}{\G(\D-2)}H^{1}_{-\f{3}{2},\f{1}{2}} \] G^+_{\D-1}
\ee
and 
\be\label{psn}
\begin{gathered}
\Psi_k(\D) = \bigg[H^{-k}_{\f{k-2}{2},\f{k}{2}}\(-H^{1}_{-\f{1}{2},-\f{1}{2}}\)^{k+1} - \f{(-1)^k}{k!}\f{\G(\D+k-2)}{\G(\D-2)} H^{0}_{-1,0}\(-H^{1}_{-\f{1}{2},-\f{1}{2}}\)\\ - \f{(-1)^k k}{(k+1)!} \f{\G(\D+k-2)}{\G(\D-3)} H^{1}_{-\f{3}{2},-\f{1}{2}}\bigg] G^+_{\D-2} 
\end{gathered}
\ee
respectively, where $k=1,2,3,\cdots, \infty$. However, it is more convenient to work with the new basis defined by 
\be
\begin{gathered}
\Omega_k(\D) = \sum_{n=1}^{k} \f{1}{(k -n)!} \f{\G(\D+k-2)}{\G(\D +n-2)}\Phi_n(\D)
\end{gathered} 
\label{nb1}
\ee
for the $\mathcal{O}({z^0\bar z^0})$ null states and similarly for the $\mathcal{O}({z^0\bar z})$ null states the new basis is defined by,
\be\label{nb2}
\Pi_k(\D) = \sum_{n=1}^{k} \f{1}{(k-n)!}\f{\G(\D+k-2)}{\G(\D+n-2)}\Psi_n(\D)
\ee
There is another set of null states, which are of the Knizhnik-Zamolodchikov type and decoupling of these null states give rise to differential equations for the scattering amplitudes \cite{Banerjee:2020zlg, Banerjee:2020vnt, Banerjee:2023rni, Hu:2021lrx, Fan:2022vbz, Fan:2022kpp, Hu:2022bpa}. We will discuss about these null states in the context of self dual gravity in section \ref{kzbz}.
Then, using these new basis \eqref{nb1} and \eqref{nb2} the OPE between two positive helicity outgoing graviton primaries with dimensions $ \D_1 $ and $ \D_2 $ of any $ w $-invariant theory can always be written as,
\be\label{0}
\begin{gathered}
G^{+}_{\D_1}(z,\bar z)G^+_{\D_2}(0,0) = -\frac{\bar z}{z} B\(\D_1 -1, \D_2 -1\) G^+_{\D_1+\D_2}(0,0) \\
 + G^{+}_{\D_1}(z,\bar z)G^+_{\D_2}(0,0)\big|_{\text{MHV at} \ \mathcal{O}(z^0\bar z^0)} + \sum_{p=1}^n B(\D_1-1 + p,\D_2-1) \  \Omega_{p}(\D_1+\D_2) \\
 + G^{+}_{\D_1}(z,\bar z)G^+_{\D_2}(0,0)\big|_{\text{MHV at} \  \mathcal{O}(z^0\bar z^1)} + \bar z \sum_{p=1}^n B(\D_1+p,\D_2-1) \ \Pi_p(\D_1+\D_2+1) + \cdots
\end{gathered} 
\ee
where $ G^{+}_{\D_1}(z,\bar z)G^+_{\D_2}(0,0)\big|_{\text{MHV at} \ \mathcal{O}(z^0\bar z^0)} $ and $ G^{+}_{\D_1}(z,\bar z)G^+_{\D_2}(0,0)\big|_{\text{MHV at} \  \mathcal{O}(z^0\bar z^1)} $ are the MHV OPEs at $\mathcal{O}({z^0\bar z^0})$ and $\mathcal{O}({z^0\bar z})$ respectively. It has been shown in \cite{Pate:2019lpp} that the leading term in $ \bar z $ is uniquely determined by the $ sl_2(R)_V $ invariance. Once the leading term is known, the subleading terms in $ \bar z $ of $ \mathcal{O}\(\f{\bar z^q}{z}\) $, $ q \geq 2 $ are determined by the $ \overline{sl_2(R)} $ invariance.

It was shown in \cite{Banerjee:2023zip}, that both the MHV null states $ \Omega_{k}(\D) $ and $ \Pi_k(\D) $ form representations of $sl_2(R)_V$. However, these representations are reducible because for any integer $ n \geq 0 $, the subspaces spanned by $ \{\Omega_{n+1}(\D), \Omega_{n+2}(\D), \cdots \} $ and $ \{\Pi_{n+1}(\D), \Pi_{n+2}(\D), \cdots \} $ form a representation of $sl_2(R)_V$. Hence we can get smaller representations spanned by the states $ \{\Omega_{1}(\D), \Omega_{2}(\D),  \cdots, \Omega_{n}(\D) \} $ and $ \{\Pi_{1}(\D), \Pi_{2}(\D), \cdots, \Pi_{n}(\D) \} $ if we set
\be \label{wi}
\begin{gathered}
\Omega_{k+1}(\D) = 0, \qquad k \geq n \geq 0, \\
\Pi_{k+1}(\D) = 0, \qquad k \geq n \geq 0.
\end{gathered}
\ee
Using the algebra \eqref{hsa}, one can also check that the null states $ \Omega_{k}(\D) $ and $ \Pi_k(\D) $ are primaries under $\overline{sl_2(R)}$. Thus the conditions \eqref{wi} are invariant under $\overline{sl_2(R)}$, hence under whole $ w $-algebra.

We have showed in section \ref{reviewalg} that, the whole tower of $ w $-currents can be generated using two sub-algebras given by 
$\overline{sl_2(R)}$ and $sl_2(R)_V$. Moreover, the conditions \eqref{wi} are also invariant under $\overline{sl_2(R)}$ and $sl_2(R)_V$, and hence under the full $ w $-algebra. Now, using these facts and the algebra \eqref{hsa}, it is not hard to show the OPE \eqref{0} is invariant under $ w $-algebra. The important point we want to emphasize about the OPE \eqref{0} is that the integer $ n $ can take any arbitrary value without breaking the $ w $-invariance. Hence, there exists a discrete infinite family of $ w $-invariant OPEs. From \eqref{0} it is already clear that $ n=0 $ gives the $ MHV $-sector. In this paper, we have shown that $ n=4 $ gives the OPE of the quantum self-dual gravity theory which is known to be $w$-invariant.

Now, the last thing we want to discuss in this section is that, the null states $ \{\Omega_{1}(\D), \\ \Omega_{2}(\D),  \cdots, \Omega_{n}(\D) \} $ are not completely independent. For a given $n$, there is another set of $\lceil{\frac{n}{2}}\rceil$\footnote{$\lceil{\frac{n}{2}}\rceil = \text{Smallest integer}$ $\ge \frac{n}{2}$.} nontrivial\footnote{There are of course the $n$ states $\{\Omega_1(\D),...,\Omega_n(\D)\}$ which transform in a representation of $sl_2(R)_V$ but, we cannot set them to zero because that will lead us again to the MHV sector.} states $\{\chi^1_n(\D),...,\chi_n^{\lceil{n/2}\rceil}(\D)\}$ defined as
\be\label{add1}
\begin{gathered}
\chi^1_n(\D) = \sum_{p=1}^n \Omega_p(\D)\\
\chi^i_n(\D) = \sum_{p=i}^n\prod_{q=i}^{2i-2}(p-q) \Omega_p(\D), \  i = 2,3,...,\lceil{\frac{n}{2}}\rceil
\end{gathered}
\ee
which transform in a representation of the $sl_2(R)_V$ as a consequence of \eqref{wi}. We can also set these states to zero 
\be\label{add2}
\chi^i_n(\D) = 0 
\ee
without violating the $sl_2(R)_V$ or $\overline{sl_2(R)}$ symmetry.

\section{Null States in Self Dual Gravity}
\label{nssd}

In this Appendix, we will derive the null states of the self dual gravity appearing at different orders of the OPE. We will first start with the OPE between two positive helicity outgoing gravitons in the self dual gravity derived in section \ref{opext}. It is given by,
\be\label{cso}
\begin{gathered}
G^{+}_{\D_4}(z_4,\bar z_4)G^+_{\D_5}(z_5,\bar z_5) = -\frac{\bar z_{45}}{z_{45}}  B\(\D_4 -1, \D_5 -1\)  G^+_{\D_4+\D_5}(z_5, \bar z_5) \\
+ B\(\D_4 -1, \D_5 -1\) \, H^1_{-\f{3}{2},\f{1}{2}} G^+_{\D_4+\D_5-1}(z_5,\bar z_5) + \sum_{k=1}^4 B(\D_4+k-1,\D_5-1)\Omega_k(\D_4+\D_5) \\
+ \bar z_{45} \[ B(\D_4-1,\D_5-1) \, G^{+}_{\D_4}(z_4,\bar z_4)G^+_{\D_5}(z_5, \bar z_5)\big|_{\text{MHV at} \  \mathcal{O}(\bar z_{45})}  \right. \\
\left. + \sum_{k=1}^4 B(\D_4+k,\D_5-1) \ \Pi_k(\D_4+\D_5+1) \] + \cdots 
\end{gathered} 
\ee
where $ G^{+}_{\D_4}(z_4,\bar z_4)G^+_{\D_5}(z_5, \bar z_5)\big|_{\text{MHV at} \  \mathcal{O}(\bar z_{45})} $ is given by \eqref{mhvzb}. We now derive the null states appearing at $ \mathcal{O}(1) $ and $ \mathcal{O}(\bar z_{45}) $.

\subsection{Null States at $ \mathcal{O}(1) $}
\label{nso1}

We can see from \eqref{cso} that at $ \mathcal{O}(1) $ the OPE truncates at $ k=4 $. Now we take the conformal soft limit $ \D_4 \to -4 $ in \eqref{cso}. In this limit, the soft descendant that appear at $ \mathcal{O}(1) $ on the LHS of \eqref{cso} is given by $ H^{-4}_{1,3}G^+_{\D_5}(z_5,\bar z_5) $. After taking the same conformal soft limits on the RHS and comparing the results we get
\be\label{om5}
\Omega_5(\D) = \sum_{j=1}^{5} \f{1}{(5 -j)!} \f{\G(\D+3)}{\G(\D +j-2)}\Phi_j(\D) = 0
\ee
where $ \Phi_j(\D) $ are given by \eqref{phn}. Thus, we see that $ \Omega_5(\D) $ is a null state of the self dual gravity. Now we will show the consistency of \eqref{om5} under $ w $-algebra. Under $ {sl_2(R)}_V $, $ \Om_5(\D) $ transforms as \eqref{nt1},\eqref{nt2},
\be\label{fs}
\begin{split}
H^{1}_{-\f{1}{2},-\f{1}{2}}\Omega_5(\D) &= - \Omega_5(\D+1)\\
H^{-1}_{\f{1}{2},\f{1}{2}}\Omega_5(\D) &= 21 \, \Omega_5(\D-1) - \f{1}{2}(\D-4)(\D-5)\Omega_5(\D-1) - 21 \, \Omega_{6}(\D-1)
\end{split} 
\ee
and $H^0_{0,0} = 2 \bar L_0$ is diagonal on these states. However, $ \Omega_6(\D-1) $ is also a null state of the theory and thus \eqref{om5} is invariant under $ {sl_2(R)}_V $. One can also check that
\be \label{omh0}
H^0_{0,1} \Omega_5(\D) = 0
\ee
Thus we see that \eqref{om5} is also invariant under $ \overline{sl_2(R)} $. Hence we conclude that \eqref{om5} is invariant under $ w $-algebra.

There is another set of null states \eqref{add1} at $ \mathcal{O}(1) $ which can be found using the commutativity property of the OPE together with the conformal soft limits. In case of self dual gravity, they are explicitly given by,
\be\label{adds1}
\begin{gathered}
\chi^1_4(\D) = \sum_{p=1}^4 \Omega_p(\D)\\
\chi^2_4(\D) = \sum_{p=3}^4 (p-2) \Omega_p(\D), 
\end{gathered}
\ee

These null states also transform under the representation of $ {sl_2(R)}_V $ and $ \overline{sl_2(R)} $ algebra and as a consequence one can set them to 0 without violating the $ w $-symmetry. The null states \eqref{adds1} play an important role in showing the invariance of the Knizhnik-Zamolodchikov type null state under $ w $-algebra which will be discussed in the next subsection.

\subsection{Null States at $ \mathcal{O}(\bar z_{45}) $: Knizhnik-Zamolodchikov Type Null State}
\label{kzbz}

KZ type null state occur at $\mathcal{O}(z^0\bar z^1)$ of the OPE. The easiest way to derive it is to use the commutativity property of the OPE and conformal soft limits together. So we start with the commutativity property of the OPE given by,
\be\label{comm}
G^+_{\D_1}(z_1, \bar z_1)G^+_{\D_2}(z_2,\bar z_2) = G^+_{\D_2}(z_2, \bar z_2)G^+_{\D_1}(z_1,\bar z_1)
\ee
Now we use the OPE \eqref{cso} in \eqref{comm}, and take the leading conformal soft limits $ \D_1 \to 1 $. Then by comparing the terms at $ \mathcal{O}(\bar z_{45}) $ we get the following KZ type equation,
\be\label{KZ}
\Xi_4(\D) = \xi(\D) + \sum_{k=1}^4 \Pi_k(\D+1) = 0
\ee
where $\xi(\D)$ is the KZ type null state in the MHV sector given by \cite{Banerjee:2020zlg}
\be
\xi(\D) = \boxed{L_{-1}} G^{+}_{\D} + H^{0}_{0,-1}H^{1}_{-\f{3}{2},\f{1}{2}}G^+_{\D-1} + H^{0}_{-1,0}G^+_{\D} + (\D-1) \, H^{1}_{-\f{3}{2},-\f{1}{2}}G^+_{\D-1}
\ee
We have used that $\chi^1_4(\D)$ is a null state in this theory to arrive at the form \eqref{KZ}. One can check that \eqref{KZ} is consistent under the actions of $\overline{sl_2(R)}$ and $sl_2(R)_V$ generators. For example, 
\be
H^{0}_{0,1}\Xi_4(\D) = 6 \, \Omega_{5}(\D) - (\D-3) \chi^1_4(\D) 
\ee
We have already shown that $\Omega_{5}(\D)$ and $\chi^1_4(\D)$ are both null states in this theory, so we get,
\be
H^{0}_{0,1}\Xi_4(\D) = 0
\ee
Therefore $\Xi_4(\D)$ is an $\overline{sl_2(R)}$ primary. \\
\\
Similarly, we have 
\be
\begin{gathered}
H^{-1}_{\f{1}{2},\f{1}{2}}\Xi_4(\D) = -\f{1}{2}(\D-2)(\D-3)\Xi_4(\D-1) \\
 - 9 \, \Pi_{5}(\D) - H^{-1}_{\f{1}{2},-\f{1}{2}} \chi^1_4(\D) - H^{0}_{0,-1} \( (\D-1)\chi^1_4(\D-1) + \chi^2_4(\D-1)\) \\
 \end{gathered} 
\ee
However, since $\Pi_{5}(\D), \chi^1_4(\Delta)$ and $\chi^2_4(\D)$ are null states in the theory, we get 
\be
H^{-1}_{\f{1}{2},\f{1}{2}}\Xi_4(\D) = -\f{1}{2}(\D-2)(\D-3)\Xi_4(\D-1)
\ee
Therefore, $\Xi_4(\D)$ transforms under a representation of the $sl_2(R)_V$ and we can consistently set it to zero without violating the $sl_2(R)_V$ symmetry. Hence, we conclude that \eqref{KZ} is indeed $w$ invariant. Decoupling of null states gives rise to differential equations which the graviton scattering amplitudes in this theory have to satisfy.

\section{Invariance of the Self-Dual OPE Under $ w $-Algebra}
\label{invsd}

In \cite{Banerjee:2023zip}, it was shown that the OPE \eqref{0} is invariant under $ w $-algebra for any arbitrary truncation in $ n $, which has been reviewed in Appendix \ref{revgs}. We have shown in section \ref{opext} that self dual OPE truncates at $ n=4 $ of the general OPE \eqref{0}. Thus, we can say that the invariance of the self dual OPE under $ w $-algebra is guaranteed. However, for the sake of completeness of this paper and for the better readability, we will repeat the same analysis here with focusing on the self dual OPE. As discussed in section \ref{reviewalg}, the whole $ w $-algebra can be derived by the combined action of $ {sl_2(R)}_V $ and $ \overline{sl_2(R)} $. Thus it is enough to show the invariance of the OPE under these two sub-algebras.
 
\subsection{$ w $-Invariance at $ \mathcal{O}(1) $}

Let us start with the $ \mathcal{O}(1) $ OPE. We write it here again for the readers convenience,
\be 
\begin{gathered}
G^{+}_{\D_1}(z,\bar z)G^+_{\D_2}(0,0) \big|_{\mathcal{O}(1)} = B(\D_1-1,\D_2-1) H^1_{-\f{3}{2},\f{1}{2}} G^+_{\D_1+\D_2-1}(0,0) \\
+ \sum_{k=1}^4 B(\D_1+k-1,\D_2-1)\Omega_k(\D_1+\D_2)\\
\end{gathered}
\label{fm_OPEO2}
\ee
We now show that it is invariant under the two subalgebras $ {sl_2(R)}_V $ and $ \overline{sl_2(R)} $.

\subsubsection{ $ {sl_2(R)}_V $ Invariance}

To show the invariance of the OPE, we need the action of the $sl_2(R)_V$ on the MHV null states $\Omega_k(\D)$ that can appear at $\mathcal{O}(1)$. These actions were computed in \cite{Banerjee:2023zip} and reviewed in Appendix \ref{mnt}. We also need the commutator algebra \eqref{hsa} along with the action of these generators on the graviton primaries given by (see Appendix \ref{wap}),
\be\label{pra}
\begin{split}
 H^{1}_{-\f{1}{2},-\f{1}{2}}G^+_{\D}(z,\bar z) &= -G^+_{\D+1}(z,\bar z)\\
 H^{-1}_{\f{1}{2},\f{1}{2}}G^+_{\D}(z,\bar z) &= -\f{1}{2}\[(\D-2)(\D-3) + 4(\D-2) \bar z \pa_{\bar z} + 3 \bar z^2 \pa_{\bar z}^2 \] G^+_{\D-1}(z,\bar z)
\end{split} 
\ee
Using Appendix \ref{mnt}, \eqref{hsa} and \eqref{pra}, it is not difficult to show that the $ \mathcal{O}(1) $ OPE \eqref{fm_OPEO2} is invariant under $ H^{1}_{-\f{1}{2},-\f{1}{2}} $ whereas the action of $ H^{-1}_{\f{1}{2},\f{1}{2}} $ on both the sides of the OPE $ \eqref{fm_OPEO2} $ gives
\be 
H^{-1}_{\f{1}{2},\f{1}{2}}\(\tn{R.H.S} \ - \ \tn{L.H.S}\) \ \tn{of} \ \eqref{fm_OPEO2} = - 12 B(\D_1+3,\D_2-1) \, \Omega_5(\D_1+\D_2-1)
\ee
However, we have already shown in Appendix \ref{nso1}, that $ \Omega_5(\D) $ is a null state of the self dual gravity appearing at $ \mathcal{O}(1) $ of the OPE and as a consequence we can set it to 0. Hence, we conclude that the $ \mathcal{O}(1) $ self dual OPE \eqref{fm_OPEO2} is invariant under the $ {sl_2(R)}_V $ algebra.

\subsubsection{ $ \overline{sl_2(R)} $ Invariance}

It was shown in \cite{Banerjee:2020zlg}, that the OPE in the MHV-sector is invariant under the action of $ H^0_{0,1} $ \footnote{$ H^0_{0,1} \sim \bar L_{1} $}. Also from \eqref{nt3}, we can see that the null states $ \Omega_k(\D) $ are annihilated by $ H^0_{0,1} $. Therefore, we can say that $ \mathcal{O}(1) $ self dual OPE \eqref{fm_OPEO2} is invariant under the $ \overline{sl_2(R)} $ algebra.

\subsection{$ w $-Invariance at $ \mathcal{O}(\bar z) $}

We now move on to showing the $ w $-invariance of the self dual OPE at  $ \mathcal{O}(\bar z) $. Let us first write down the $ \mathcal{O}(\bar z) $ OPE \eqref{ozm} again,
\be\label{ozi}
\begin{gathered}
G^{+}_{\D_1}(z,\bar z)G^+_{\D_2}(0,0)\big|_{\mathcal{O}(\bar z)} = B(\D_1-1,\D_2-1) \, G^{+}_{\D_1}(z,\bar z)G^+_{\D_2}(0, 0)\big|_{\text{MHV at} \  \mathcal{O}(\bar z_{45})} \\
+ \sum_{k=1}^4 B(\D_1+k,\D_2-1) \ \Pi_k(\D_1+\D_2+1)
\end{gathered} 
\ee
From the previous subsection, it is clear that the $ w $-invariance of the OPE at $ \mathcal{O}(\bar z) $ is guaranteed to follow if we can show that it is invariant under the two subalgebras $ {sl_2(R)}_V $ and $ \overline{sl_2(R)} $. Among the generators of these two subalgebras, we only show the invariance of the OPE \eqref{ozi} under the actions of $ H^{-1}_{\f{1}{2},\f{1}{2}} $ and $ H^0_{0,1} $. This is mainly because the invariance of the OPE \eqref{ozi} under the rest of the generators are fairly easy to show. By applying $ H^{-1}_{\f{1}{2},\f{1}{2}} $ on both sides of the OPE \eqref{ozi} we get,
\be
H^{-1}_{\f{1}{2},\f{1}{2}}\(\tn{R.H.S} \ - \ \tn{L.H.S}\) \ \tn{of} \ \eqref{ozi} = - 9 B(\D_1+4,\D_2-1) \, \Pi_5(\D_1+\D_2) 
\ee
and for $ H^0_{0,1} $ we have 
\be
H^{(0)}_{0,1}\(\tn{R.H.S} \ - \ \tn{L.H.S}\) \ \tn{of} \ \eqref{ozi} = 6 B(\D_4+4,\D_5-1) \, \Omega_5(\D_1+\D_2) 
\ee
However, from Appendix \ref{nssd}, we know that both $ \Pi_5(\D) $ and $ \Omega_5(\D) $ are the null states of the self dual gravity appearing at $ \mathcal{O}(\bar z) $ and $ \mathcal{O}(1) $ respectively. Thus, we conclude that the $ \mathcal{O}(\bar z)  $ OPE in self dual gravity is also invariant under $ {sl_2(R)}_V $ and $ \overline{sl_2(R)} $, and hence under the whole $ w $-algebra.



\begin{thebibliography}{99}

\bibitem{Strominger:2017zoo}
A.~Strominger,
``Lectures on the Infrared Structure of Gravity and Gauge Theory,''
[arXiv:1703.05448 [hep-th]].




 \bibitem{Pasterski:2016qvg} 
  S.~Pasterski, S.~H.~Shao and A.~Strominger,
  ``Flat Space Amplitudes and Conformal Symmetry of the Celestial Sphere,''
  Phys.\ Rev.\ D {\bf 96}, no. 6, 065026 (2017)
  doi:10.1103/PhysRevD.96.065026
  [arXiv:1701.00049 [hep-th]].

  \bibitem{Pasterski:2017kqt} 
  S.~Pasterski and S.~H.~Shao,
  ``Conformal basis for flat space amplitudes,''
  Phys.\ Rev.\ D {\bf 96}, no. 6, 065022 (2017)
  doi:10.1103/PhysRevD.96.065022
  [arXiv:1705.01027 [hep-th]].  
  
   \bibitem{Banerjee:2018gce} 
  S.~Banerjee,
  ``Null Infinity and Unitary Representation of The Poincare Group,''
  JHEP {\bf 1901}, 205 (2019)
  doi:10.1007/JHEP01(2019)205
  [arXiv:1801.10171 [hep-th]].
  
\bibitem{Banerjee:2019prz}
S.~Banerjee, S.~Ghosh, P.~Pandey and A.~P.~Saha,
``Modified celestial amplitude in Einstein gravity,''
JHEP \textbf{03} (2020), 125
doi:10.1007/JHEP03(2020)125
[arXiv:1909.03075 [hep-th]].
  

\bibitem{Ooguri:1991fp}
H.~Ooguri and C.~Vafa,
``Geometry of N=2 strings,''
Nucl. Phys. B \textbf{361} (1991), 469-518
doi:10.1016/0550-3213(91)90270-8

\bibitem{Chalmers:1996rq}
G.~Chalmers and W.~Siegel,
``The Selfdual sector of QCD amplitudes,''
Phys. Rev. D \textbf{54} (1996), 7628-7633
doi:10.1103/PhysRevD.54.7628
[arXiv:hep-th/9606061 [hep-th]].


\bibitem{Bern:1998xc}
Z.~Bern, L.~J.~Dixon, M.~Perelstein and J.~S.~Rozowsky,
``One loop n point helicity amplitudes in (selfdual) gravity,''
Phys. Lett. B \textbf{444} (1998), 273-283
doi:10.1016/S0370-2693(98)01397-5
[arXiv:hep-th/9809160 [hep-th]].

\bibitem{Bern:1998sv}
Z.~Bern, L.~J.~Dixon, M.~Perelstein and J.~S.~Rozowsky,
``Multileg one loop gravity amplitudes from gauge theory,''
Nucl. Phys. B \textbf{546} (1999), 423-479
doi:10.1016/S0550-3213(99)00029-2
[arXiv:hep-th/9811140 [hep-th]].

\bibitem{Krasnov:2021sf}
K.~Krasnov, 
``Self-Dual Gravity,''
Class. Quant. Grav. \textbf{34} (2017) 095001 [1610.01457].

\bibitem{Penrose:1968pr}
R.~Penrose, ``Twistor quantization and curved space-time,'' Int. J. Theor. Phys. \textbf{1} (1968) 61.

\bibitem{Penrose:1976pr}
R.~Penrose, ``Nonlinear Gravitons and Curved Twistor Theory,'' Gen. Rel. Grav. \textbf{7} (1976) 31.

\bibitem{Boyer:1985pn}
C.~P.~Boyer and J.~F.~Plebanski, ``An infinite hierarchy of conservation laws and nonlinear superposition principles for selfdual Einstein spaces,''
 J. Math. Phys. 26 (1985) 229.

   \bibitem{Sachs:1962zza} 
  R.~Sachs,
  ``Asymptotic symmetries in gravitational theory,''
  Phys.\ Rev.\  {\bf 128}, 2851 (1962).
  doi:10.1103/PhysRev.128.2851

  H.~Bondi, M.~G.~J.~van der Burg and A.~W.~K.~Metzner,
 ``Gravitational waves in general relativity. 7. Waves from axisymmetric isolated systems,''
  Proc.\ Roy.\ Soc.\ Lond.\ A {\bf 269}, 21 (1962).
  doi:10.1098/rspa.1962.0161
  
  R.~K.~Sachs,
  ``Gravitational waves in general relativity. 8. Waves in asymptotically flat space-times,''
  Proc.\ Roy.\ Soc.\ Lond.\ A {\bf 270}, 103 (1962).
  doi:10.1098/rspa.1962.0206


\bibitem{Strominger:2013jfa} 
  A.~Strominger,
  ``On BMS Invariance of Gravitational Scattering,''
  JHEP {\bf 1407}, 152 (2014)
  doi:10.1007/JHEP07(2014)152
  [arXiv:1312.2229 [hep-th]].
  
 \bibitem{He} 
  T.~He, V.~Lysov, P.~Mitra and A.~Strominger,
  ``BMS supertranslations and Weinberg's soft graviton theorem,''
  JHEP {\bf 1505}, 151 (2015)
  doi:10.1007/JHEP05(2015)151
  [arXiv:1401.7026 [hep-th]].
  
  \bibitem{Strominger:2014pwa} 
 A.~Strominger and A.~Zhiboedov,
 ``Gravitational Memory, BMS Supertranslations and Soft Theorems,''
 JHEP {\bf 1601}, 086 (2016)
doi:10.1007/JHEP01(2016)086
[arXiv:1411.5745 [hep-th]].

   \bibitem{Barnich:2009se} 
  G.~Barnich and C.~Troessaert,
  ``Symmetries of asymptotically flat 4 dimensional spacetimes at null infinity revisited,''
 Phys.\ Rev.\ Lett.\  {\bf 105}, 111103 (2010)
  doi:10.1103/PhysRevLett.105.111103
  [arXiv:0909.2617 [gr-qc]]. 
  
  G.~Barnich and C.~Troessaert,
  ``Supertranslations call for superrotations,''
  PoS CNCFG {\bf 2010}, 010 (2010)
  [Ann.\ U.\ Craiova Phys.\  {\bf 21}, S11 (2011)]
  [arXiv:1102.4632 [gr-qc]].
  
  \bibitem{Kapec:2016jld}
  D.~Kapec, P.~Mitra, A.~M.~Raclariu and A.~Strominger,
  ``2D Stress Tensor for 4D Gravity,''
  Phys.\ Rev.\ Lett.\  {\bf 119}, no. 12, 121601 (2017)
  doi:10.1103/PhysRevLett.119.121601
  [arXiv:1609.00282 [hep-th]].
      
  \bibitem{Kapec:2014opa} 
  D.~Kapec, V.~Lysov, S.~Pasterski and A.~Strominger,
  ``Semiclassical Virasoro symmetry of the quantum gravity $ \mathcal{S}$-matrix,''
  JHEP {\bf 1408}, 058 (2014)
  doi:10.1007/JHEP08(2014)058
  [arXiv:1406.3312 [hep-th]].
    
  \bibitem{He:2017fsb} 
  T.~He, D.~Kapec, A.~M.~Raclariu and A.~Strominger,
  ``Loop-Corrected Virasoro Symmetry of 4D Quantum Gravity,''
  JHEP {\bf 1708}, 050 (2017)
  doi:10.1007/JHEP08(2017)050
  [arXiv:1701.00496 [hep-th]].
  
\bibitem{Banerjee:2022wht}
S.~Banerjee and S.~Pasterski,
``Revisiting the shadow stress tensor in celestial CFT,''
JHEP \textbf{04} (2023), 118
doi:10.1007/JHEP04(2023)118
[arXiv:2212.00257 [hep-th]].

\bibitem{Donnay:2021wrk}
L.~Donnay and R.~Ruzziconi,
``BMS flux algebra in celestial holography,''
JHEP \textbf{11}, 040 (2021)
doi:10.1007/JHEP11(2021)040
[arXiv:2108.11969 [hep-th]].

\bibitem{Donnay:2022hkf}
L.~Donnay, K.~Nguyen and R.~Ruzziconi,
``Loop-corrected subleading soft theorem and the celestial stress tensor,''
JHEP \textbf{09}, 063 (2022)
doi:10.1007/JHEP09(2022)063
[arXiv:2205.11477 [hep-th]].

  
\bibitem{Donnay:2020guq}
L.~Donnay, S.~Pasterski and A.~Puhm,
``Asymptotic Symmetries and Celestial CFT,''
JHEP \textbf{09} (2020), 176
doi:10.1007/JHEP09(2020)176
[arXiv:2005.08990 [hep-th]].

  
  \bibitem{Stieberger:2018onx} 
  S.~Stieberger and T.~R.~Taylor,
  ``Symmetries of Celestial Amplitudes,''
  Phys.\ Lett.\ B {\bf 793}, 141 (2019)
  doi:10.1016/j.physletb.2019.03.063
  [arXiv:1812.01080 [hep-th]].
  
  \bibitem{Banerjee:2020kaa}
S.~Banerjee, S.~Ghosh and R.~Gonzo,
``BMS symmetry of celestial OPE,''
JHEP \textbf{04}, 130 (2020)
doi:10.1007/JHEP04(2020)130
[arXiv:2002.00975 [hep-th]].
  



  
\bibitem{Banerjee:2020zlg}
S.~Banerjee, S.~Ghosh and P.~Paul,
``MHV graviton scattering amplitudes and current algebra on the celestial sphere,''
JHEP \textbf{02} (2021), 176
doi:10.1007/JHEP02(2021)176
[arXiv:2008.04330 [hep-th]].

\bibitem{Banerjee:2021cly}
S.~Banerjee, S.~Ghosh and S.~S.~Samal,
``Subsubleading soft graviton symmetry and MHV graviton scattering amplitudes,''
JHEP \textbf{08} (2021), 067
doi:10.1007/JHEP08(2021)067
[arXiv:2104.02546 [hep-th]].


\bibitem{Gupta:2021cwo}
N.~Gupta, P.~Paul and N.~V.~Suryanarayana,
`` $\widehat{sl_2}$ Symmetry of ${\mathbb R}^{1,3}$ Gravity,''
Phys. Rev. D \textbf{108} (2023) no.8, 086029
doi:10.1103/PhysRevD.108.086029
[arXiv:2109.06857 [hep-th]].

\bibitem{Guevara:2021abz}
A.~Guevara, E.~Himwich, M.~Pate and A.~Strominger,
``Holographic symmetry algebras for gauge theory and gravity,''
JHEP \textbf{11} (2021), 152
doi:10.1007/JHEP11(2021)152
[arXiv:2103.03961 [hep-th]].
 

\bibitem{Strominger:2021lvk}
A.~Strominger,
``w(1+infinity) and the Celestial Sphere,''
[arXiv:2105.14346 [hep-th]]. 

A.~Strominger,
``$w_{1+\infty}$ Algebra and the Celestial Sphere: Infinite Towers of Soft Graviton, Photon, and Gluon Symmetries,''
Phys. Rev. Lett. \textbf{127}, no.22, 221601 (2021)
doi:10.1103/PhysRevLett.127.221601

\bibitem{Himwich:2021dau}
E.~Himwich, M.~Pate and K.~Singh,
``Celestial operator product expansions and w$_{1+\infty}$ symmetry for all spins,''
JHEP \textbf{01}, 080 (2022)
doi:10.1007/JHEP01(2022)080
[arXiv:2108.07763 [hep-th]].

\bibitem{Melton:2022fsf}
W.~Melton, S.~A.~Narayanan and A.~Strominger,
``Deforming soft algebras for gauge theory,''
JHEP \textbf{03}, 233 (2023)
doi:10.1007/JHEP03(2023)233
[arXiv:2212.08643 [hep-th]].

\bibitem{Ball:2021tmb}
A.~Ball, S.~A.~Narayanan, J.~Salzer and A.~Strominger,
``Perturbatively exact w$_{1+ \infty}$ asymptotic symmetry of quantum self-dual gravity,''
JHEP \textbf{01}, 114 (2022)
doi:10.1007/JHEP01(2022)114
[arXiv:2111.10392 [hep-th]].



\bibitem{Adamo:2021lrv}
T.~Adamo, L.~Mason and A.~Sharma,
``Celestial $w_{1+\infty}$ Symmetries from Twistor Space,''
SIGMA \textbf{18}, 016 (2022)
doi:10.3842/SIGMA.2022.016
[arXiv:2110.06066 [hep-th]].

\bibitem{Costello:2022wso}
K.~Costello and N.~M.~Paquette,
``Celestial holography meets twisted holography: 4d amplitudes from chiral correlators,''
JHEP \textbf{10}, 193 (2022)
doi:10.1007/JHEP10(2022)193
[arXiv:2201.02595 [hep-th]].

\bibitem{Costello:2022upu}
K.~Costello and N.~M.~Paquette,
``Associativity of One-Loop Corrections to the Celestial Operator Product Expansion,''
Phys. Rev. Lett. \textbf{129}, no.23, 231604 (2022)
doi:10.1103/PhysRevLett.129.231604
[arXiv:2204.05301 [hep-th]].

\bibitem{Costello:2023vyy}
K.~J.~Costello,
``Bootstrapping two-loop QCD amplitudes,''
[arXiv:2302.00770 [hep-th]].

\bibitem{Garner:2023izn}
N.~Garner and N.~M.~Paquette,
``Twistorial monopoles \& chiral algebras,''
JHEP \textbf{08}, 088 (2023)
doi:10.1007/JHEP08(2023)088
[arXiv:2305.00049 [hep-th]].

\bibitem{Stieberger:2022zyk}
S.~Stieberger, T.~R.~Taylor and B.~Zhu,
``Celestial Liouville theory for Yang-Mills amplitudes,''
Phys. Lett. B \textbf{836}, 137588 (2023)
doi:10.1016/j.physletb.2022.137588
[arXiv:2209.02724 [hep-th]].

\bibitem{Stieberger:2023fju}
S.~Stieberger, T.~R.~Taylor and B.~Zhu,
``Yang-Mills as a Liouville theory,''
Phys. Lett. B \textbf{846}, 138229 (2023)
doi:10.1016/j.physletb.2023.138229
[arXiv:2308.09741 [hep-th]].

\bibitem{Ball:2023qim}
A.~Ball, M.~Spradlin, A.~Yelleshpur Srikant and A.~Volovich,
``Supersymmetry and the Celestial Jacobi Identity,''
[arXiv:2311.01364 [hep-th]].

\bibitem{Banerjee:2019aoy}
S.~Banerjee, P.~Pandey and P.~Paul,
``Conformal properties of soft operators: Use of null states,''
Phys. Rev. D \textbf{101}, no.10, 106014 (2020)
doi:10.1103/PhysRevD.101.106014
[arXiv:1902.02309 [hep-th]].

\bibitem{Banerjee:2019tam}
S.~Banerjee and P.~Pandey,
``Conformal properties of soft-operators. Part II. Use of null-states,''
JHEP \textbf{02}, 067 (2020)
doi:10.1007/JHEP02(2020)067
[arXiv:1906.01650 [hep-th]].

\bibitem{Pasterski:2021fjn}
S.~Pasterski, A.~Puhm and E.~Trevisani,
``Celestial diamonds: conformal multiplets in celestial CFT,''
JHEP \textbf{11}, 072 (2021)
doi:10.1007/JHEP11(2021)072
[arXiv:2105.03516 [hep-th]].

\bibitem{Pano:2023slc}
Y.~Pano, A.~Puhm and E.~Trevisani,
``Symmetries in Celestial CFT$_{d}$,''
JHEP \textbf{07}, 076 (2023)
doi:10.1007/JHEP07(2023)076
[arXiv:2302.10222 [hep-th]].

\bibitem{Banerjee:2023zip}
S.~Banerjee, H.~Kulkarni and P.~Paul,
``An infinite family of w$_{1+\infty}$ invariant theories on the celestial sphere,''
JHEP \textbf{05} (2023), 063
doi:10.1007/JHEP05(2023)063
[arXiv:2301.13225 [hep-th]].


\bibitem{Banerjee:2020vnt}
S.~Banerjee and S.~Ghosh,
``MHV gluon scattering amplitudes from celestial current algebras,''
JHEP \textbf{10} (2021), 111
doi:10.1007/JHEP10(2021)111
[arXiv:2011.00017 [hep-th]].

\bibitem{Ebert:2020nqf}
S.~Ebert, A.~Sharma and D.~Wang,
``Descendants in celestial CFT and emergent multi-collinear factorization,''
JHEP \textbf{03}, 030 (2021)
doi:10.1007/JHEP03(2021)030
[arXiv:2009.07881 [hep-th]].

\bibitem{Banerjee:2023rni}
S.~Banerjee, R.~Mandal, A.~Manu and P.~Paul,
``MHV gluon scattering in the massive scalar background and celestial OPE,''
JHEP \textbf{10} (2023), 007
doi:10.1007/JHEP10(2023)007
[arXiv:2302.10245 [hep-th]].

\bibitem{Adamo:2022wjo}
T.~Adamo, W.~Bu, E.~Casali and A.~Sharma,
``All-order celestial OPE in the MHV sector,''
JHEP \textbf{03}, 252 (2023)
doi:10.1007/JHEP03(2023)252
[arXiv:2211.17124 [hep-th]]

\bibitem{Ren:2023trv}
L.~Ren, A.~Schreiber, A.~Sharma and D.~Wang,
``All-order celestial OPE from on-shell recursion,''
JHEP \textbf{10}, 080 (2023)
doi:10.1007/JHEP10(2023)080
[arXiv:2305.11851 [hep-th]].

\bibitem{Hu:2021lrx}
Y.~Hu, L.~Ren, A.~Y.~Srikant and A.~Volovich,
``Celestial dual superconformal symmetry, MHV amplitudes and differential equations,''
JHEP \textbf{12}, 171 (2021)
doi:10.1007/JHEP12(2021)171
[arXiv:2106.16111 [hep-th]].

\bibitem{Fan:2022vbz}
W.~Fan, A.~Fotopoulos, S.~Stieberger, T.~R.~Taylor and B.~Zhu,
``Elements of celestial conformal field theory,''
JHEP \textbf{08}, 213 (2022)
doi:10.1007/JHEP08(2022)213
[arXiv:2202.08288 [hep-th]].

\bibitem{Fan:2022kpp}
W.~Fan, A.~Fotopoulos, S.~Stieberger, T.~R.~Taylor and B.~Zhu,
``Celestial Yang-Mills amplitudes and D = 4 conformal blocks,''
JHEP \textbf{09}, 182 (2022)
doi:10.1007/JHEP09(2022)182
[arXiv:2206.08979 [hep-th]].

\bibitem{Bhardwaj:2022anh}
R.~Bhardwaj, L.~Lippstreu, L.~Ren, M.~Spradlin, A.~Yelleshpur Srikant and A.~Volovich,
``Loop-level gluon OPEs in celestial holography,''
JHEP \textbf{11}, 171 (2022)
doi:10.1007/JHEP11(2022)171
[arXiv:2208.14416 [hep-th]].

\bibitem{Krishna:2023ukw}
H.~Krishna,
``Celestial gluon and graviton OPE at loop level,''
[arXiv:2310.16687 [hep-th]].


\bibitem{Hu:2022bpa}
Y.~Hu and S.~Pasterski,
``Celestial recursion,''
JHEP \textbf{01} (2023), 151
doi:10.1007/JHEP01(2023)151
[arXiv:2208.11635 [hep-th]].


  

  
  


 
 
 

    

  
  
  
\bibitem{Pate:2019lpp}
M.~Pate, A.~M.~Raclariu, A.~Strominger and E.~Y.~Yuan,
``Celestial operator products of gluons and gravitons,''
Rev. Math. Phys. \textbf{33} (2021) no.09, 2140003
doi:10.1142/S0129055X21400031
[arXiv:1910.07424 [hep-th]].
  
\bibitem{Atanasov:2021cje}
A.~Atanasov, W.~Melton, A.~M.~Raclariu and A.~Strominger,
``Conformal block expansion in celestial CFT,''
Phys. Rev. D \textbf{104} (2021) no.12, 126033
doi:10.1103/PhysRevD.104.126033
[arXiv:2104.13432 [hep-th]].

  
  \bibitem{Ghosh:2022net}
S.~Ghosh, P.~Raman and A.~Sinha,
``Celestial insights into the S-matrix bootstrap,''
JHEP \textbf{08}, 216 (2022)
doi:10.1007/JHEP08(2022)216
[arXiv:2204.07617 [hep-th]]


  
  \bibitem{Donnay:2018neh} 
  L.~Donnay, A.~Puhm and A.~Strominger,
  ``Conformally Soft Photons and Gravitons,''
  JHEP {\bf 1901}, 184 (2019)
  doi:10.1007/JHEP01(2019)184
  [arXiv:1810.05219 [hep-th]].
      
\bibitem{Pate:2019mfs}
M.~Pate, A.~M.~Raclariu and A.~Strominger,
``Conformally Soft Theorem in Gauge Theory,''
Phys. Rev. D \textbf{100} (2019) no.8, 085017
doi:10.1103/PhysRevD.100.085017
[arXiv:1904.10831 [hep-th]].
  
  \bibitem{Fan:2019emx} 
  W.~Fan, A.~Fotopoulos and T.~R.~Taylor,
  ``Soft Limits of Yang-Mills Amplitudes and Conformal Correlators,''
  JHEP {\bf 1905}, 121 (2019)
  doi:10.1007/JHEP05(2019)121
  [arXiv:1903.01676 [hep-th]]. 
  
\bibitem{Nandan:2019jas}
D.~Nandan, A.~Schreiber, A.~Volovich and M.~Zlotnikov,
``Celestial Amplitudes: Conformal Partial Waves and Soft Limits,''
JHEP \textbf{10} (2019), 018
doi:10.1007/JHEP10(2019)018
[arXiv:1904.10940 [hep-th]].
     
\bibitem{Adamo:2019ipt}
T.~Adamo, L.~Mason and A.~Sharma,
``Celestial amplitudes and conformal soft theorems,''
Class. Quant. Grav. \textbf{36} (2019) no.20, 205018
doi:10.1088/1361-6382/ab42ce
[arXiv:1905.09224 [hep-th]].
    
\bibitem{Puhm:2019zbl}
A.~Puhm,
``Conformally Soft Theorem in Gravity,''
JHEP \textbf{09} (2020), 130
doi:10.1007/JHEP09(2020)130
[arXiv:1905.09799 [hep-th]].
 
 \bibitem{Guevara:2019ypd} 
  A.~Guevara,
  ``Notes on Conformal Soft Theorems and Recursion Relations in Gravity,''
  arXiv:1906.07810 [hep-th].
  
\bibitem{Cotler:2023qwh}
J.~Cotler, N.~Miller and A.~Strominger,
``An integer basis for celestial amplitudes,''
JHEP \textbf{08}, 192 (2023)
doi:10.1007/JHEP08(2023)192
[arXiv:2302.04905 [hep-th]].
  
\bibitem{Freidel:2022skz}
L.~Freidel, D.~Pranzetti and A.~M.~Raclariu,
``A discrete basis for celestial holography,''
[arXiv:2212.12469 [hep-th]].
 
\end{thebibliography}
\end{document}